\documentclass{aa}  
\usepackage{bm}
\usepackage{graphicx}
\usepackage{comment}
\usepackage{txfonts}
\usepackage[breaklinks=true, draft]{hyperref}
\begin{document} 

\title{Structure, kinematics, and ages of the young stellar populations \\ in the Orion region}

\author{E. Zari \inst{1}, A. G. A. Brown \inst{1}, and P.T. de Zeeuw \inst{1}
}
\institute{
{1} Leiden Observatory, Leiden University, Niels Bohrweg 2, 2333 CA Leiden, the Netherlands; \\
}

\abstract{
We present a study of the three dimensional structure, kinematics, and age distribution of the Orion OB association, based on the second data release of the \textit{Gaia} satellite (\textit{Gaia} DR2). Our goal is to obtain a complete picture of the star formation history of the Orion complex and to relate our findings to theories of sequential and triggered star formation.  We select the Orion population with simple photometric criteria, and we construct a three dimensional map in galactic Cartesian coordinates to study the physical arrangement of the stellar clusters in the Orion region. The map shows structures that extend for roughly $150 \, \mathrm{pc}$ along the line of sight, divided in multiple sub-clusters.  We separate different groups by using the density based clustering algorithm DBSCAN. 
We study the kinematic properties of all the groups found by DBSCAN first by inspecting their proper motion distribution, and then by applying a kinematic modelling code based on an  iterative maximum likelihood approach, which we use to derive their mean velocity, velocity dispersion and isotropic expansion. 
By using an isochrone fitting procedure we provide ages and extinction values for all the groups. We confirm the presence of an old population ($\sim 15$ Myr) towards the 25 Ori region, and we find that groups with ages of $12-15 \, \mathrm{Myr}$ are present also towards the Belt region. We notice the presence of a population of $\sim 10$ Myr also in front of the Orion A molecular cloud. 
Our findings suggest that star formation in Orion does not follow a simple sequential scenario, but instead consists of multiple events, which caused kinematic and physical sub-structure. To fully explain the detailed sequence of events, specific simulations and further radial velocity data are needed. 
}


\titlerunning{Orion DR2}

\maketitle
%
\section{Introduction}
The tendency of O and B type stars to loosely cluster in the sky was recognized at the beginning of the 20th century by the pioneering studies 
summarized in \cite{Blaauw1964}. At the end of the last century, the data of the \textit{Hipparcos} satellite allowed \cite{deZeeuw1999, deBruijne1999, Hoogerwerf1999}, and many others, to characterise the stellar content and the kinematic properties of nearby OB associations.  
Due to the fact that their members  are widely dispersed over the sky, OB associations have been long considered as expanding remnants of young star clusters \citep{Brown1999, Lada2003}.
The classical explanation for this is that star clusters are formed embedded within molecular clouds, where the gravitational potential of both the stars and the gas holds them together. When feedback disperses the gas left over from star formation, the cluster becomes supervirial and will expand and disperse, thus being visible for a short time as an OB association.
While many observations support this model \citep[and references therein]{Lada2003}, it has been difficult to test whether OB associations are indeed expanding. 
\cite{Wright2016} and  \cite{Wright2018} studied the kinematics of the Cygnus OB2 and Scorpius-Centaurus associations respectively, and concluded that they were not formed by the disruption of individual star clusters. \cite{Wright2018} further concluded that Sco-Cen was likely born highly sub-structured, with multiple small-scale star formation events contributing to the overall OB association, and not as a single, monolithic burst of clustered star formation.
These conclusions can be related to the fact that the distribution of young stars within their parental molecular clouds is fractal and hierarchical, and follows the filamentary structures of the dense gas 
both spatially \citep{Gutermuth2008} and kinematically \citep{Hacar2016}.
Clusters then form where filaments overlap \citep{Myers2009, Schneider2012, Hacar2016, Hacar2017}: their formation might be due to higher column densities or to the merging of filaments that have already formed stars.
OB associations would therefore constitute the final stage of this star formation mechanism, as, while they are slowly dispersing in the field, they still keep memory of the parental gas sub-structure where they originated.

At a distance of $ \sim 380 \, \mathrm{pc}$  \citep{Zari2017}, the Orion star forming region is the nearest site of active high mass star formation.
It is a benchmark for studying all stages and modes of star formation  \citep[e.g.,][]{Brown1994, Jeffries2006, Bally2008, Briceno2008, Muench2008, DaRio2014,  Getman2014, DaRio2016,  Hacar2016, Kubiak2017, Fang2017, Kounkel2017b}, in addition to the effect of star formation processes on the surrounding interstellar medium \citep{Ochsendorf2015, Schlafly2015, Soler2018}.
\cite{Zari2017} used \textit{Gaia} DR1 \citep{Brown2016, Prusti2016} to study the density distribution of the young, non-embedded stellar population in the sky, and obtained a first picture of the star formation history of the Orion region in terms of the various star formation episodes, their duration, and their effects on the surrounding interstellar medium. Even though proper motions where available for the TGAS \citep{Michalik2015} sub-set of \textit{Gaia} DR1, they were not accurate enough to perform a precise kinematic analysis. Proper motions in Orion are indeed small as stars move on average radially away from the Sun. Furthermore, to derive the ages of the stellar populations, a single distance value was considered ($d \sim 380 \, \mathrm{pc}$), as parallax uncertainties were too large to resolve the spatial configuration of the groups that were identified. 
By combining the data of the second release of the \textit{Gaia} satellite \citep[hereafter \textit{Gaia} DR2][]{Brown2018} and APOGEE-2, \cite{Kounkel2018} study the entire Orion complex, providing a classification of the stellar population in five groups, and an analysis of their ages and kinematics.
\cite{Kos2018} use \textit{Gaia} DR2 parameters supplemented with radial velocities from the GALAH and APOGEE surveys to perform a clustering analysis towards the 25 Ori cluster region, and find that one cluster is significantly older ($21 \pm 2 \, \mathrm{Myr}$) compared to the rest of the region.
\cite{Grossschedl2018} investigate the 3D shape and orientation of the Orion A molecular cloud by analysing the distances of mid-infrared selected young stellar objects, and find that the cloud is elongated and oriented towards the galactic plane, and presents two different components, one dense and star forming, and one $\sim 75 \, \mathrm{pc}$ long, more diffuse and star-formation quiet.

In this work, we use \textit{Gaia} DR2 to study the three dimensional (3D) structure of the Orion OB association, we model the kinematics of the sub-groups that constitute it and we give estimates of their ages, to obtain a complete picture of the star formation history of the region and to put it in the broader context of the theories of sequential and triggered star formation.
In Section 2 we present the data and describe how we select the young stellar population in Orion.
In Section 3 we study its 3D configuration in Cartesian galactic coordinates, and we isolate young groups by making use of the DBSCAN clustering algorithm.  In Section 4 we perform the kinematic analysis by using a maximum likelihood approach. In Section 5 we derive ages and extinctions of all the groups resulting from the analysis of Section 4. In Section 6 we discuss our findings. The conclusions of this work are summarised in Section 7.

\section{Data}\label{sec:2}
\begin{figure*}[h]
    \includegraphics[width = \hsize, ]{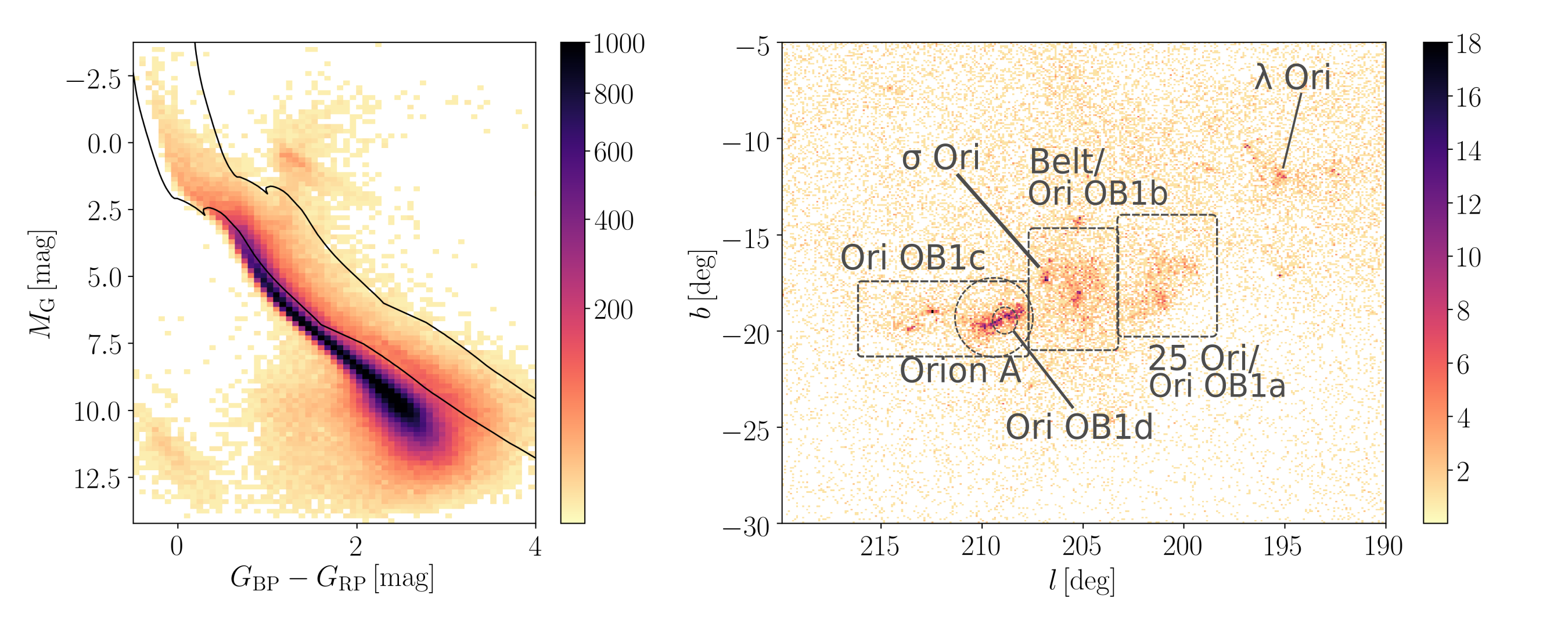}
    \caption{Observed colour-magnitude diagram (left) and  sky distribution (right) of the sources selected in Section 2. The solid black lines in the left panel show the isochrones defined in Eq. \ref{eq:eq6}, which are used to select the young stellar population in Orion. }
    \label{fig:fig1}
\end{figure*}
Following \cite{Zari2017}, we select the sources with coordinates
\begin{align}
    190^{\circ} < l < 220^{\circ}, \qquad
    -30^{\circ} < b < -5^{\circ},
\end{align}
and we restrict our sample to the sources with $1.5 < \varpi < 5.0 \, \mathrm{mas}$.
Since the Orion population moves mostly radially away from the Sun, we consider only stars with small proper motions:
\begin{equation}
    |\mu_{\alpha*}|  < 10 \,  \mathrm{mas \, yr^{-1}},  \qquad
    |\mu_{\delta}|   < 10 \, \mathrm{mas \, yr^{-1}}.
\end{equation}
We derive distances by inverting parallaxes, $d = 1000/\varpi \, \mathrm{pc}$ thus we restrict our sample to sources with $\varpi/\sigma_{\varpi} > 5.0$, following the recommendations in  \cite{BailerJones2015}.
The effect of this cut is to exclude sources at faint magnitudes ($G > 20 \, \mathrm{mag}$), but it does not introduce significant biases in e.g., the determination of distances to the clusters or the study of their 3D configuration.

\subsection{Obtaining a 'clean' sample}
We apply the following cuts on the photometric and astrometric quality, based on \cite{Lindegren2018} complemented by the information contained on the \textit{Gaia} known issues page (\url{https://www.cosmos.esa.int/web/gaia/dr2-known-issues}):
\begin{itemize}
    \item The Renormalized Unit Weight Error (RUWE) is defined as:
    \begin{equation}
        \mathrm{RUWE} = \sqrt{\chi^2 / (N - 5)}/u_0(C, G)
    \end{equation}
where: 
\begin{itemize}
    \item $\chi^2$ is the astrometric goodness-of-fit in the AL direction (\texttt{astrometric\_chi2\_al});
    \item $N$ is the number of good observations AL (\texttt{astrometric\_n\_good\_obs\_al});
    \item $u_0(C, G)$ is an empirical normalization factor, which is a function of $C = G_{\mathrm{BP}}-G_{\mathrm{RP}}$ and $G$.
\end{itemize}
and we select all the sources with RUWE < 1.4, following the slides by Lindegren et al. (see \url{https://www.cosmos.esa.int/web/gaia/dr2-known-issues}).  This cut seeks to remove sources with spurious parallaxes or proper motions.

\item We use the flux excess ratio: 
\begin{equation}
    E = (I_{\mathrm{BP}} + I_{\mathrm{\mathrm{RP}}})/I_G,
\end{equation}
where $I_{\mathrm{X}}$ is the photometric flux in band $X$, to exclude sources with possible issues in the $\mathrm{BP}$ and $\mathrm{\mathrm{RP}}$ photometry,  affecting in particular faint sources in crowded areas. We apply Eq. C.2 in \cite{Lindegren2018}, which we report here for clarity:
\begin{equation}
    1.0  + 0.015(G_{\mathrm{BP}} - G_{\mathrm{\mathrm{RP}}})^2 < E < 1.3 + 0.06(G_{\mathrm{BP}} - G_{\mathrm{\mathrm{RP}}})^2.
\end{equation}
\end{itemize}

\noindent
\cite{Evans2018} and \cite{Arenou2018} mention that \textit{Gaia} DR2 photometry is affected by some systematic errors.  \cite{Evans2018} and \cite{Maiz2018} propose corrections to mitigate these effects. We apply these corrections and we report them here for clarity:

\begin{itemize}
    \item $2 \le G \le 6$ mag: \\
    $G_{\mathrm{corr}} = -0.047344 + 1.16405G - 0.046799G^2 + 0.0035015G^3$ \\
    
    \item $2 \le G \le 4$ mag: \\
    $G_{\mathrm{BP, corr}} = G_{\mathrm{BP}} -2.0384 + 0.95282G-0.11018G^2$ \\
    \item $2 \le G \le 3.5$ mag: \\
    $G_{\mathrm{\mathrm{RP}, corr}} = G_{\mathrm{\mathrm{RP}}} -13.946 + 14.239G_{\mathrm{RP}} -4.23G_{\mathrm{RP}}^2 + 0.4532G_{\mathrm{RP}}^3$\\
    
    \item $6 \le G \le 16$ mag:\\
    $G_{\mathrm{corr}} = G - 0.0032(G - 6.0) $\\
    
    \item $G > 16$ mag: \\
    $G_{\mathrm{corr}} = G  - 0.032 $
\end{itemize}
In the rest of the paper we use the corrected $G$, $G_{\mathrm{BP}}$, and $G_{\mathrm{RP}}$ magnitudes without using the subscript "\textit{corr}".

\subsection{Selecting the young stellar popoulation}
Figure \ref{fig:fig1} (left) shows the  $M_G$ vs. $G_{BP} - G_{\mathrm{RP}}$ colour-magnitude diagram of the 'clean' sample obtained in Section 2.1. Although faint, the pre-main sequence and the upper main sequence, indicating the presence of the young population in the region, are visible, and can be used to guide the selection of the young stellar populations towards Orion.
\newline
To select young stars, we use the PARSEC isochrones \citep{Bressan2012, Tang2014, Chen2014}  with $A_V  = 0.3 \, \mathrm{mag}$ and age $\tau = 10 \, \mathrm{Myr}$ to define the following region in the  $M_G$ vs. $G_{BP} - G_{\mathrm{RP}}$ colour-magnitude diagram (solid black lines in Fig. \ref{fig:fig1}):
\begin{align}\label{eq:eq6}
  G_{BP} - G_{\mathrm{RP}} - 0.2 & \le  M_G  \\ \nonumber
  G_{BP} - G_{\mathrm{RP}} + 0.5 & \ge  M_G - 0.8
\end{align}
We choose $A_V = 0.3 \, \mathrm{mag}$ following \cite{Zari2017}.
The distribution in the sky of the sources selected in this fashion is shown in Fig. \ref{fig:fig1} (right). The regions in which we divide the field are also indicated, together with the sub-groups in which the Orion OB1 association is classically split: Orion OB1a, OB1b, OB1c, and OB1d. The same groups identified in \cite{Zari2017} and \cite{Kounkel2018} are visible, which confirms the correctness of the selection. 

\noindent
In Section 4 we focus on the kinematics of the Orion population. To complement the \textit{Gaia} DR2 radial velocities we cross-matched our sources with the APOGEE DR14 catalogue \citep{APOGEE2018}. The APOGEE synthetic heliocentric velocities (\texttt{SYNTHVHELIO\_AVG}, an average of the
individual measured RVs using spectra cross-correlations with single best-match synthetic spectrum) were used. 

\section{3D distribution and identification of clusters}\label{sec:3}
We first study the three-dimensional (3D) distribution of sources using a similar approach as in \cite{Zari2018}. In summary, we:
\begin{enumerate}
\item compute galactic Cartesian coordinates for all the sources, $x_g, y_g, z_g$;
\item define a volume, V = (800, 800, 350), centred in the Sun, and we divide it in $3 \times 3 \times 3$ pc cubes;
\item compute the number of sources in each cube;
\item compute the source density $D(x_g, y_g, z_g)$ by smoothing the distribution with a Gaussian filter, with width $w = 2 \, \mathrm{pc}$;
\item normalise the density distribution from 0 to 1 by applying the sigmoidal logistic function:
\begin{equation}
    f(D) = \frac{L}{1 + e^{-\kappa(D- D_0)}} - 1, 
\end{equation}
with $L = 2$, $\kappa = 4 \, \mathrm{pc}$, and $D_0 = 0$.
\end{enumerate}

\begin{figure*}
    \centering
    \includegraphics[width =\hsize]{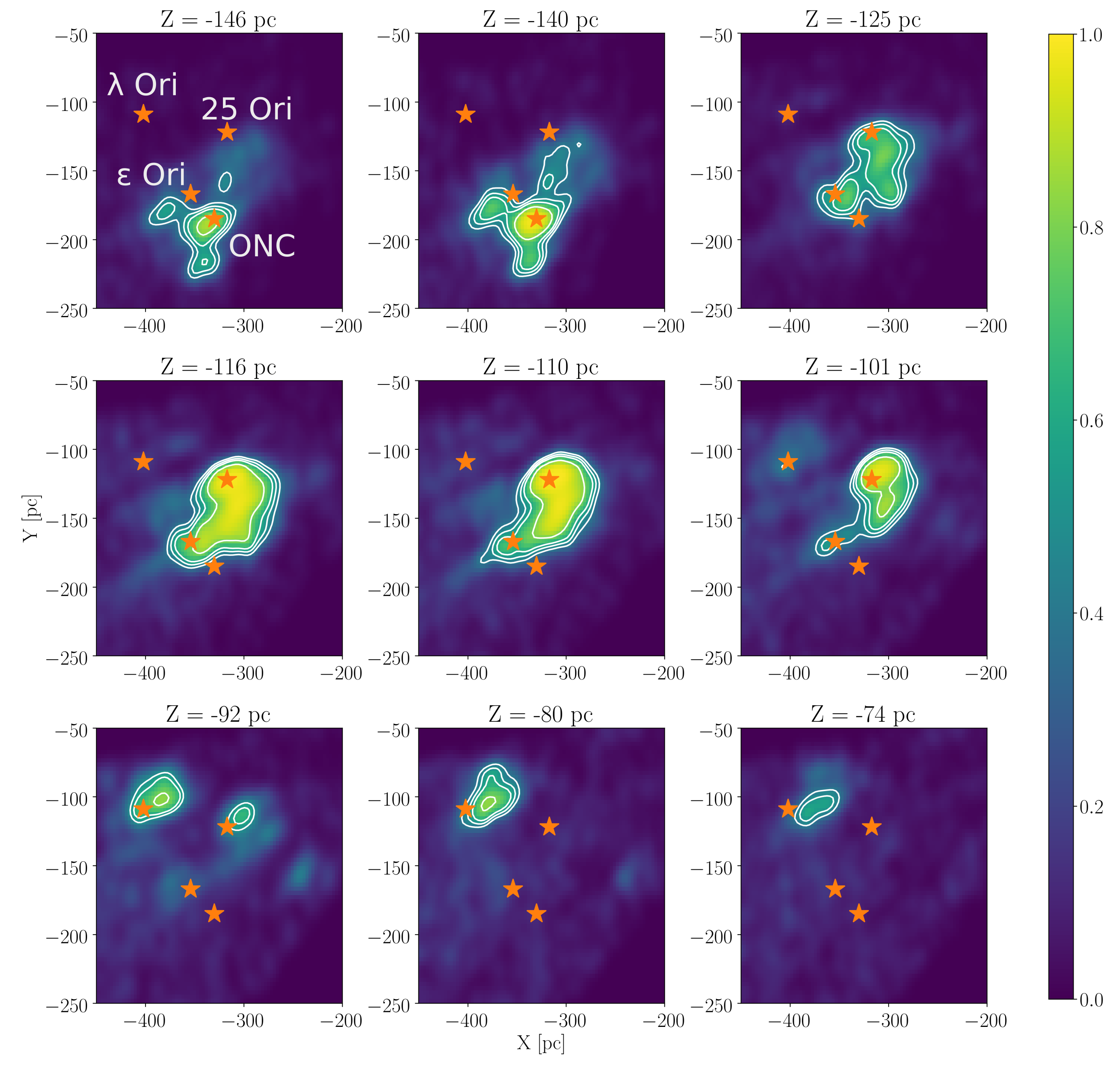}
    \caption{Density distribution of the sources in Orion for different $Z$ values. The orange stars indicate the positions of (from top to bottom in each panel): $\lambda$ Ori, 25 Ori, $\epsilon$ Ori, and the Orion Nebula cluster. The white solid  contours represent the 0.4, 0.5, 0.6 and 0.8 density levels (note that the density is normalised to have values from 0 to 1). The Sun is located at $(X, Y)  = (0, 0)$. }
    \label{fig:fig2}
\end{figure*}

\noindent
Fig. \ref{fig:fig2} shows  the density distribution of sources $f(D)$ on the galactic plane for different values of $z_g$. Different density enhancements are visible, corresponding to well known-clusters.
The first and second panel show stars in the Orion A molecular cloud. The Orion Nebula Cluster (ONC) corresponds to the most prominent density enhancement. The third panel is particularly interesting because it clearly shows the presence of a foreground population to the ONC, confirming the conclusions by \cite{Bouy2014}.  Some clusters corresponding to the Belt region also become visible, although the bulk of the population is located between $Z = -116 \, \mathrm{pc}$ and $Z = -101 \, \mathrm{pc}$.
The last three panels mainly show the $\lambda$ Ori cluster. At $Z = -92 \, \mathrm{pc}$ the Northern elongation of the 25 Ori group is visible.
The density distribution looks elongated towards the line of sight: this is an effect of the parallax errors. The parallax error distribution is peaked at $\sigma_{\varpi} = 0.046 \, \mathrm{mas}$, but presents a long tail towards larger values (the $84th$ percentile is $0.11 \, \mathrm{mas}$). 

\begin{figure*}
    \centering
    \includegraphics[width =\hsize]{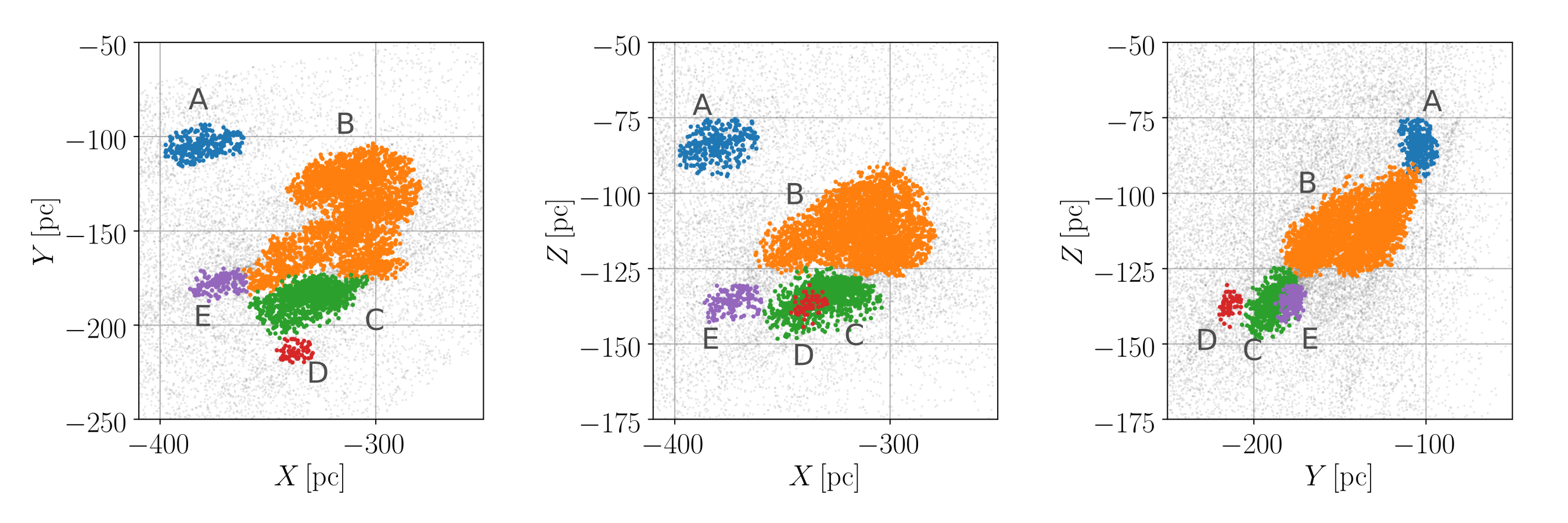}
    \includegraphics[width =\hsize]{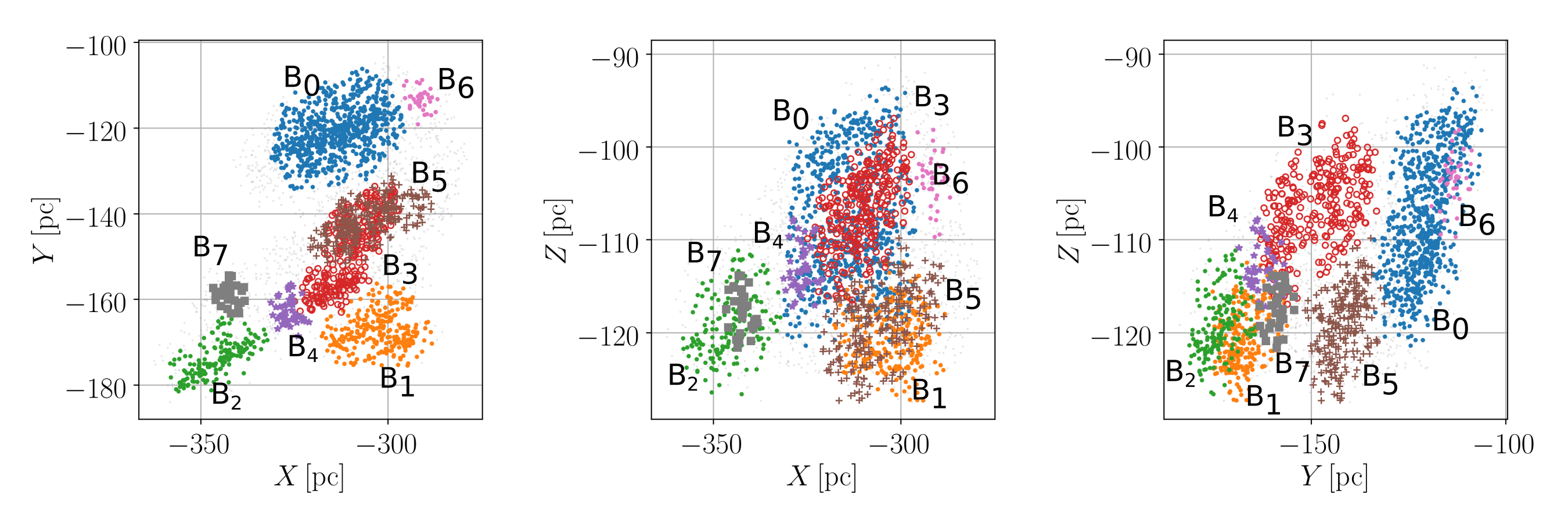}
    \caption{Distribution of the groups identified with the first (top) and second (bottom) iteration of DBSCAN in the planes $(X,Y)$, $(X,Z)$, and $(Y,Z)$.}
    \label{fig:fig3}
\end{figure*}

\begin{figure}
    \centering
    \includegraphics[width =\hsize]{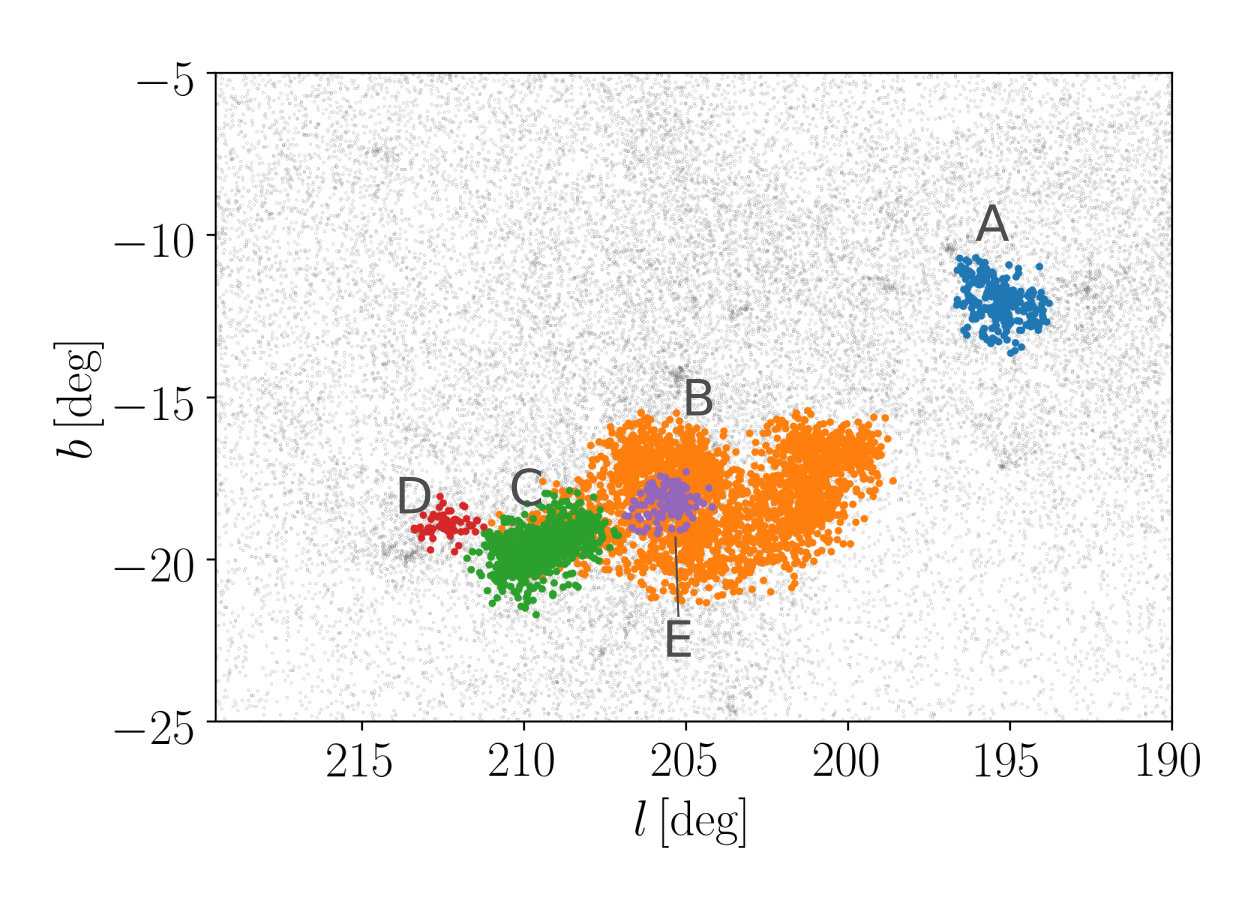}
    \includegraphics[width =\hsize]{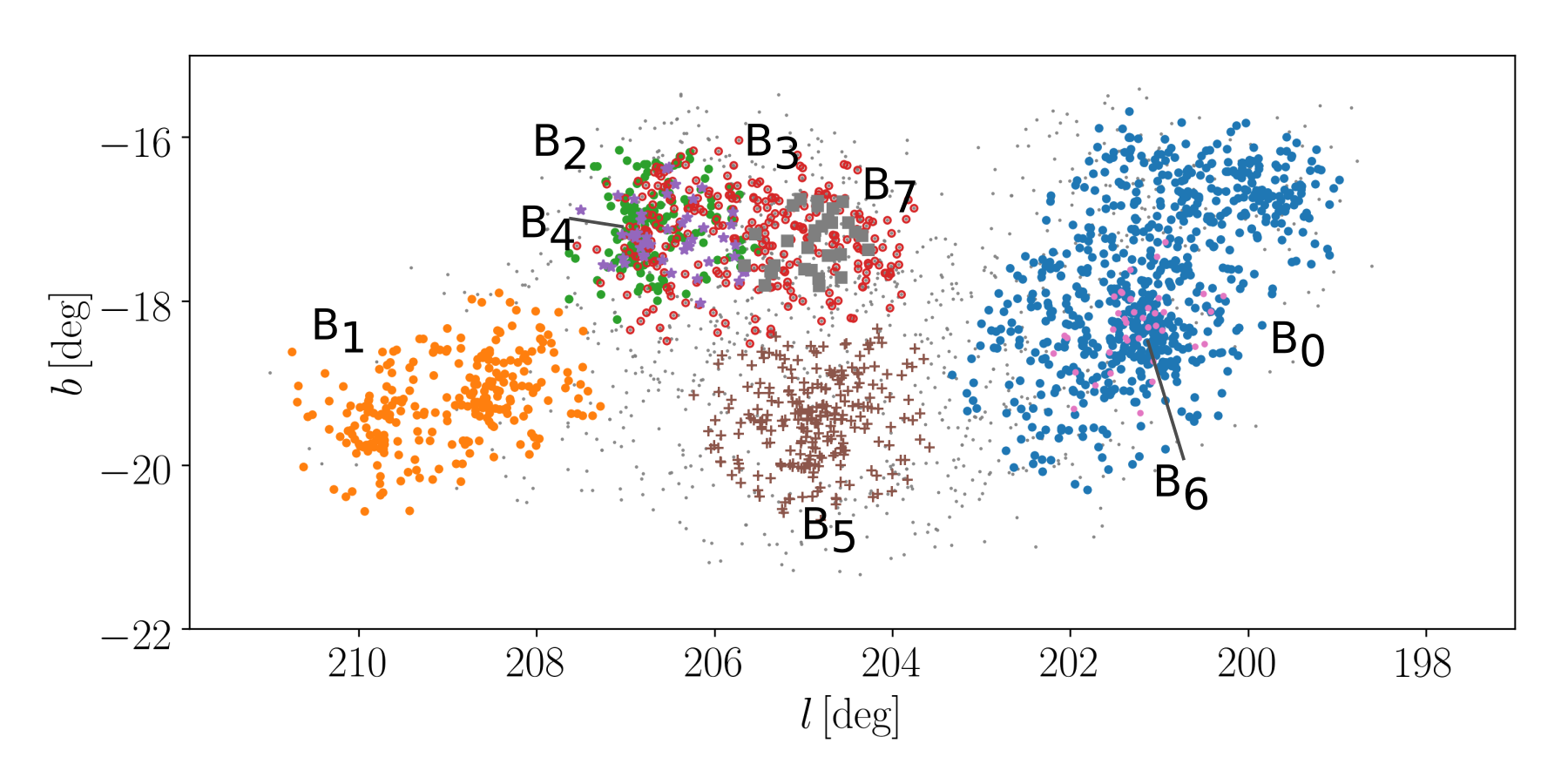}
    \caption{Sky distribution of the groups identified with the first (top) and second (bottom) iteration of DBSCAN. The colours correspond to those in Fig. \ref{fig:fig3}.}
    \label{fig:fig4}
\end{figure}

\noindent
To isolate the members of each cluster, we first consider only the sources within the density level $f(D) = 0.5$ of the 3D map shown in Fig. \ref{fig:fig2}. This value is arbitrary and aims at selecting the densest regions of the maps.
The clusters are then separated by using the DBSCAN algorithm \footnote{We use the \texttt{scikit-learn} implementation of the algorithm \citep{scikit-learn}}.  
As described  e.g. by \cite{Price-Jones2019}, DBSCAN is a density-based clustering algorithm that views clusters as areas of high density separated by areas of low density in space, without requiring any prior assumption on the number of groups present. There are two parameters to the algorithm, \texttt{min\_samples} and \texttt{eps}, which define the density of the clusters. Higher \texttt{min\_samples} or lower \texttt{eps} values indicate higher densities necessary to form a cluster.
\newline
Clusters in Orion have different sizes and numbers of members, and therefore different densities: for this reason  we need to apply the clustering algorithm twice.
The first time we use \texttt{min\_samples}= 50 and \texttt{eps}= 7 pc to isolate the main structures, shown in Figs. \ref{fig:fig3} and \ref{fig:fig4} (top), obtaining five groups. The group that encompasses 25 Ori, the Belt region and the Orion A foreground can be visibly divided in sub-groups. Thus we apply DBSCAN only to this group with different paramenters: we find that \texttt{min\_samples}= 30 and \texttt{eps}= 5 pc are the best values to separate all the sub-clusters (see Figs. \ref{fig:fig3} and \ref{fig:fig4}, bottom).
\newline
This method has the drawback of excluding stars that might be related to the star formation events in Orion, but are more dispersed than the rest of the population in 3D space (but could still be compact in proper motion space). This is further discussed in Section 6.

\section{Kinematics}\label{sec:4}
In this section we study the kinematics of the groups selected in the previous section. We use an iterative maximum likelihood approach to determine a) the average motion of the groups, b) their velocity dispersion, and c) (where possible) the presence of a linear expansion term. We use the method proposed by \cite{Lindegren2000} and applied in \cite{Reino2018} and \cite{Bravi2018}, adding however a term to take into account a potential expansion of the cluster from its centre.
The method is summarised in Section \ref{sec:4.1}, tested in Appendix A, and the results are presented in Section 4.2.
Note that here we use ICRS coordinates, which we differentiate from galactic coordinates by adding the subscript '$I$' when needed.
 
\subsection{Method}\label{sec:4.1}
Our method extends the maximum-likelihood method developed by \citet[][L00]{Lindegren2000} by adding measured radial velocities \citep[see][]{Reino2018} and by including a linear expansion term in the cluster velocity model.
Following L00, we assume that the members of a cluster share the same three-dimensional space motion with a small isotropic dispersion term. \cite{Reino2018} extended L00's method by:
\begin{itemize}
    \item adding measured radial velocity, whenever available, as a fourth observable, besides trigonometric parallax and proper motion;
    \item making a transition from the $\chi^2$ statistic used in L00, and denoted $g$, to a $p$ value or $1-\mathrm{CDF}($g$,\mathrm{DOF})$ as a goodness-of-fit statistic;
    \item using  a  mixed  three-  and  four-dimensional  likelihood function so that both stars with and without known radial velocity can be treated simultaneously.
\end{itemize}
Following L00, we include a linear expansion term in the cluster velocity model by writing the expected space velocity of a single star at position $\bm{b}_i$ as:
\begin{equation}
\bm{u}_i = \bm{v}_0 + \bm{T}(\bm{b}_i - \bm{b}_0),
\end{equation}
where  $\bm{b}_0$ is an arbitrary reference position, namely the point where the local velocity $\bm{u}(\bm{b})$ assumes the status of 'centroid' velocity $\bm{v}_0$. The coordinates of $\bm{b}_0$ are therefore fixed in advance.
The matrix $\bm{T}$ is simply a diagonal matrix of the form:
$$
\bm{T} = \begin{bmatrix}
    \kappa & 0 & 0  \\
    0 & \kappa & 0  \\
    0 & 0 & \kappa \\
 \end{bmatrix}
$$
An expanding cluster will have $\kappa > 0$, from which an expansion age, $\tau = 1/(\gamma \kappa) \, \mathrm{[Myr]}$ can be 
derived \citep[$\gamma$ is a conversion factor of 1.0227 $\mathrm{pc \, Myr^{-1} \, km^{-1} \, s,}$ see e.g.][]{Wright2018}.
\newline
The method is applied to the members of the clusters identified in Section 3.
These clusters still  contain  'outliers', i.e. real non-members, or members which have (slightly) discrepant astrometry  (and/or  radial  velocities)  as  a  result  of  unrecognised  multiplicity, them escaping from the cluster, etc. Such outliers can be found, after maximising the likelihood function, by computing the $p$ value (associated with a particular $g$ value) for each star in the solution (Eq. 19 in L00). 
The largest outlier is removed from the sample and a new maximum likelihood solution is determined, until all $g$ values are acceptably small ($g_i \le g_{lim}$ or $p_i \ge p_{lim}$). The stopping criterion is the same as in \cite{Reino2018}, i.e. that associated to a significance level $p_{lim} = 0.0027$. As noted in \cite{Reino2018}, if one stops too early, real outliers will be left and the best-fit velocity dispersion will remain too high. On the contrary, one can keep on iterating and removing outliers until just two stars with very similar three-dimensional motions are left, severely underestimating the velocity dispersion.
Astrometric data only can not distinguish between expansion or contraction of a cluster from a change in $\bm{v}_0$ (see L00). Therefore when the fraction of measured radial velocities is lower than  the 20\% we do not estimate the expansion coefficient $\kappa$ (implicitly assuming $\kappa = 0$). The threshold is conservative for certain groups, but the derived parameters are robust for all the groups.

\subsection{Results}
The results of the kinematic modelling code are give in Table \ref{tab:table4}. 
Being quite isolated with respect to the rest of the population, the $\lambda$ Ori group (group A) is easy to identify and separate from the others, therefore the results do not require any specific clarification. This is not the case for the groups with $199^{\circ} < l < 216^{\circ}$. We comment on the results for these groups by dividing them in three 'regions' according to their sky distribution: the 25 Ori region, the Belt region, and the Orion A region.
\newline

\subsubsection{25 Ori}
\begin{figure*}
    \centering
    \includegraphics[width = 0.49\hsize]{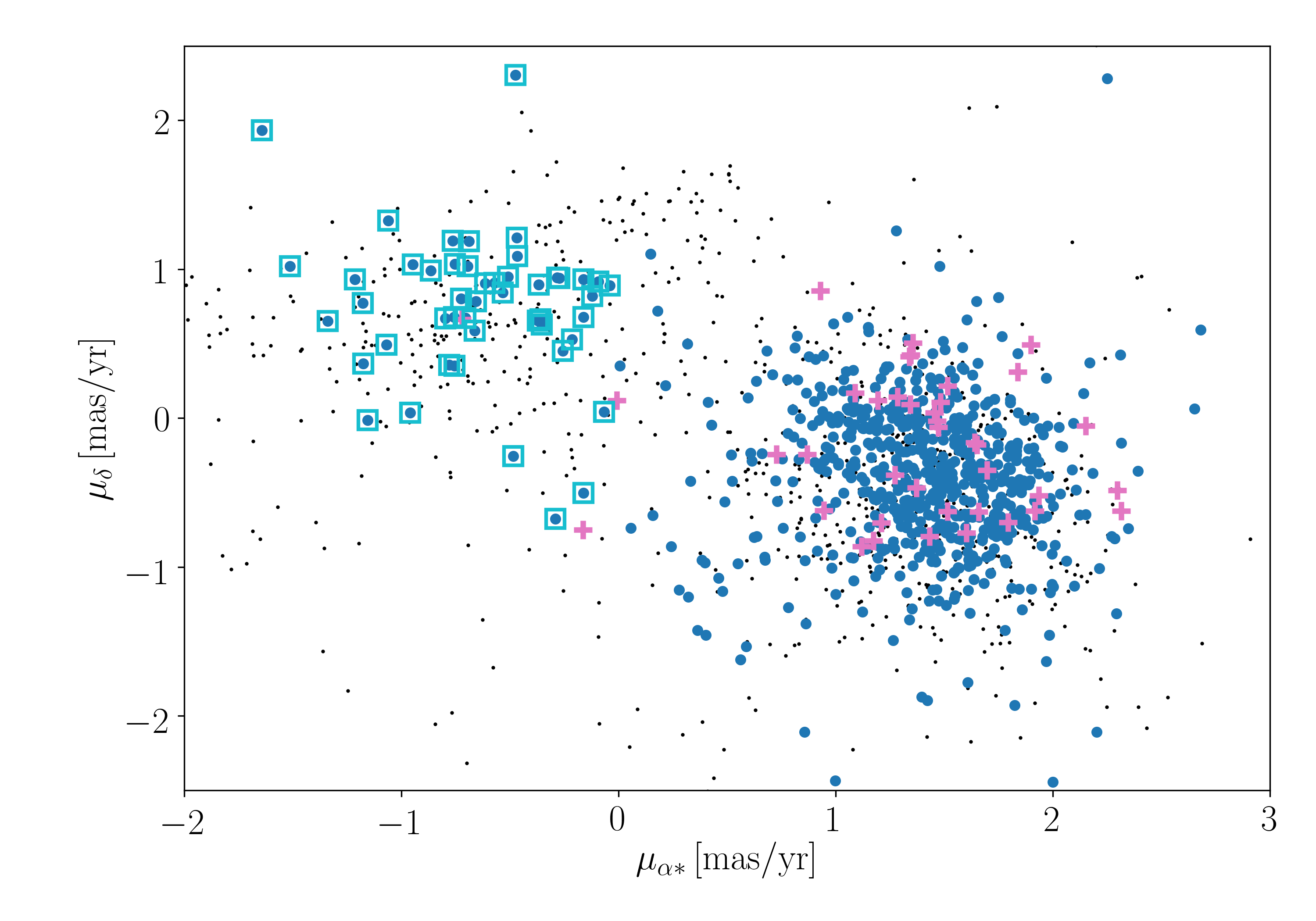}
    \includegraphics[width = 0.49\hsize]{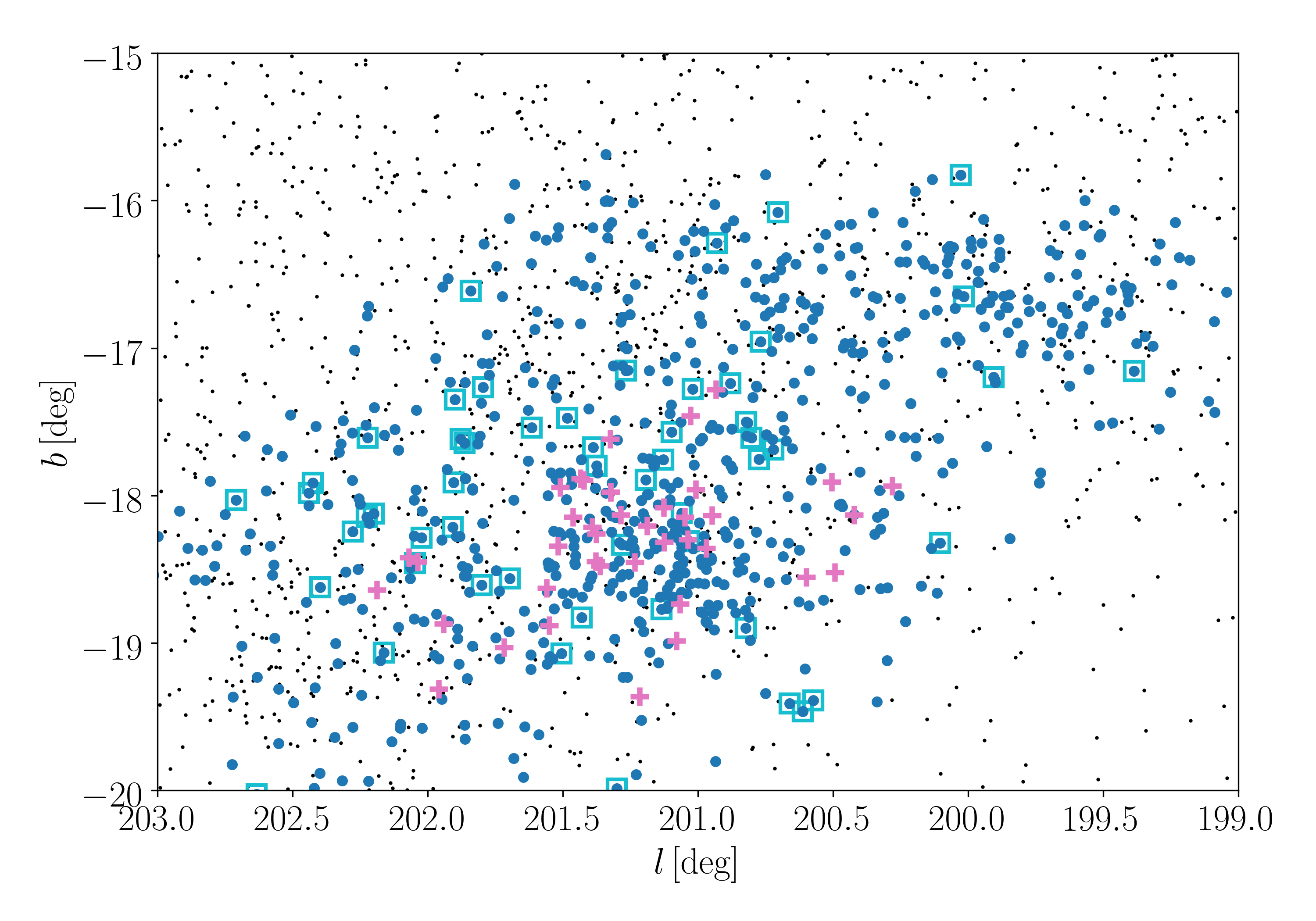}  
    \caption{Left: proper motion diagram of the stars in the 25 Ori region defined in the text (black dots), of the stars belonging to group $B_0$ (blue dots), and of the stars belonging to group $B_{0, b}$ (light blue empty squares) and $B_6$ (pink crosses).
    Right: sky distribution in galactic coordinates of group $B_{0}$, $B_{0,b}$, and $B_6$. The colours and symbols are the same as on the left.
    }
    \label{fig:fig9}
\end{figure*}

We define the 25 Ori region as:
\begin{equation}
     199^{\circ} < l < 203^{\circ} \qquad
     -20^{\circ} < b < -15^{\circ}, 
    \end{equation}
which corresponds to the groups $B_0$ and $B_6$ identified by DBSCAN.
The proper motions of the sources in the region (black dots in  Fig. \ref{fig:fig9}, left) separate in two clumps. This was shown also by \cite{Kos2018}, who however apply a different classification scheme to separate the clusters in the region. The separation is also visible when considering the proper motion diagram of group $B_0$ (blue dots in Fig. \ref{fig:fig9}, left).
The number of sources is lower because the DBSCAN algorithm favours the high density groups (so when the density drops under a certain level the stars are considered as 'noise stars' and not classified as members of any cluster).
\newline
We considered the sources selected by DBSCAN, and we isolated the second group ($B_{0, b}$, light blue squares in Fig. \ref{fig:fig9}, left) by applying the following cuts in proper motion space:
\begin{equation}\label{eq:eq9}
     \mu_{\alpha*}  < 0 \, \mathrm{mas \, yr^{-1}} \qquad
      \mu_{\delta}  > -1 \, \mathrm{mas \, yr^{-1}}. 
\end{equation}
We applied separately the kinematic modelling code to the two groups. The results are reported in Table \ref{tab:table4}. Note that we also run the kinematic modelling code considering all the sources in the region, after separating the two groups using the same criteria of Eq. \ref{eq:eq9}. The estimated parameters are consistent.
The sky distribution of the sources of group $B_0$ and $B_{0, b}$ is shown in Fig. \ref{fig:fig9} (right panel).
While group $B_0$'s distribution shows a clump towards 25 Ori, and the Northern elongation reported by e.g. \cite{Lombardi2017} and \cite{Briceno2019},  group $B_{0, b}$'s sources are scattered in the field and do not show any clear concentration. Together with the findings by \cite{Kos2018} in terms of ages (see also Section 5), this points to the conclusion that group $B_{0,b}$ is slowly dispersing in the galactic field.
Here we are limiting our samples to the 25 Ori region, but in principle members of the $B_{0,b}$ group could be found spread over a larger area of the sky (and 3D space). 
\newline
Group $B_6$ consists only of 30 members, none of which has a measured radial velocity, therefore we decided not apply the kinematic modelling code. The parallax distribution suggests that $B_6$ is closer to the Sun than group $B_0$, while the proper motion distribution does not show any difference with respect to group $B_0$. We suspect that group $B_6$ coincides with a small over-density of sources within group $B_0$, which gets classified as a separate group because of a local density drop. We ran the kinematic modelling code for groups $B_0$ and $B_6$ together: the estimated parameters are consistent with those found for group $B_0$ only, which supports our hypothesis.

\subsubsection{Belt}
\begin{figure*}
    \centering
    \includegraphics[width =\hsize]{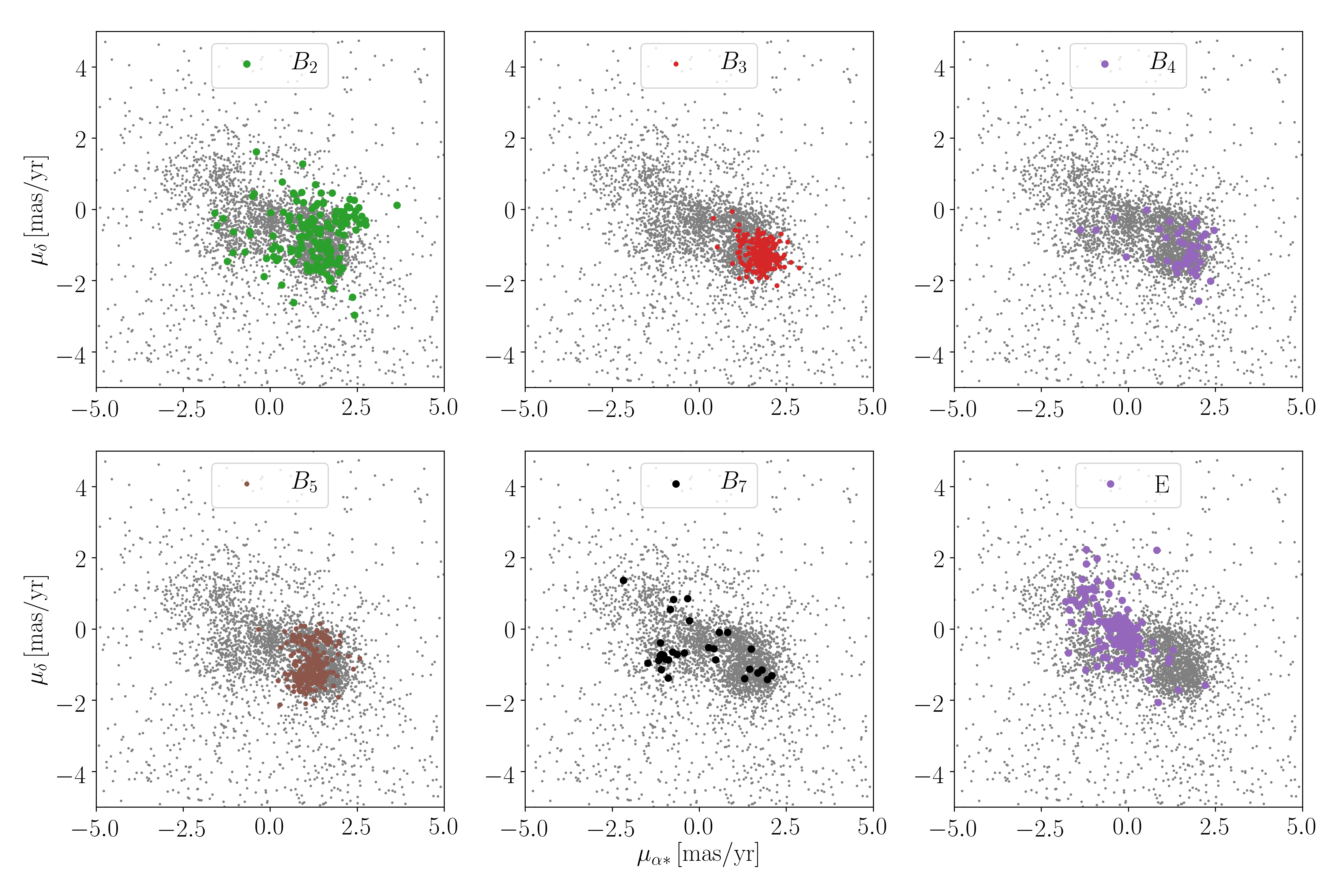}
    \caption{Proper motion diagram for all the stars in the Belt region (grey dots). Each panel corresponds to one of the groups identified by DBSCAN (the colours are the same as in Figs. \ref{fig:fig3}  and \ref{fig:fig4}) except for group $B_7$, which is indicated by black dots for representation purposes.}
    \label{fig:fig8}
\end{figure*}
Many of the clusters identified by DBSCAN ($B_2, B_3, B_4, B_5, B_7$ and E) are located in the Sky towards the Belt region.  Fig. \ref{fig:fig8} shows the proper motion diagram for the Belt region defined as 
\begin{equation}
    203^{\circ}  < l < 207.5^{\circ} \qquad 
    -21^{\circ}  < b < -13^{\circ}.
\end{equation}
Proper motions in the Belt region present a high degree of sub-structure, indicating that the Belt hosts groups with different kinematic properties.
\begin{figure*}
    \centering
    \includegraphics[width =\hsize]{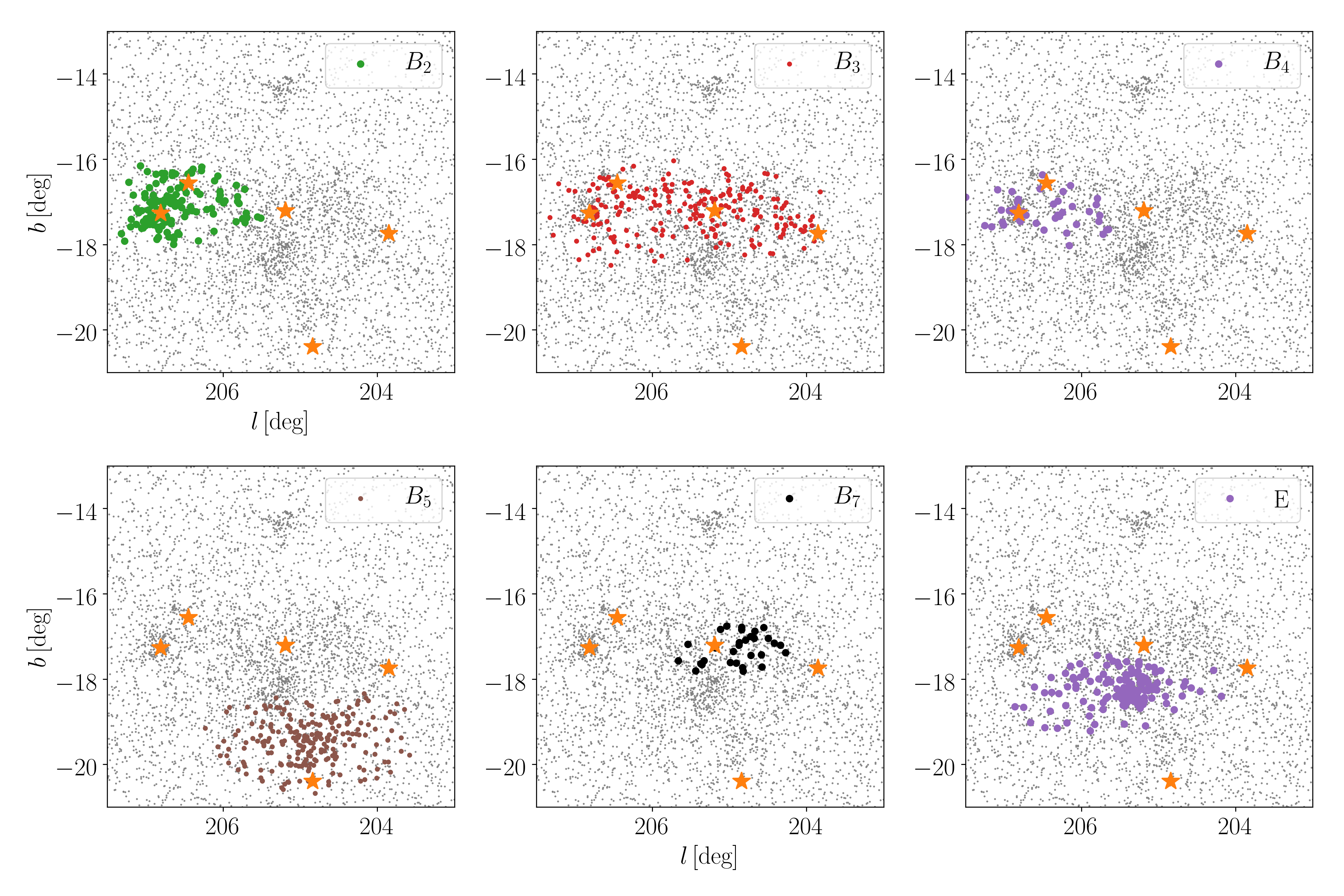}
    \caption{Distribution in the sky of the stars in the Belt region. Each panel corresponds to one group identified by DBSCAN. The colour-coding is the same as in Fig. \ref{fig:fig8}. The orange stars correspond to, from left to right: $\sigma$ Ori, $\zeta$ Ori, $\epsilon$ Ori, $\eta$ Ori, $\delta$ Ori.}
    \label{fig:fig11}
\end{figure*}
\begin{itemize}
\item Groups $B_2 $ and $B_4$ are mostly located towards the  $\sigma$ Ori cluster (see Fig. \ref{fig:fig11}) and $\zeta$ Ori. Group $B_3$'s members are spread towards $\epsilon$ Ori and $\delta$ Ori. The parameters estimated by the kinematic modelling code suggest that $B_2$ and $B_4$ have compatible $v_{y, I}$ values, which are significantly different from those of group $B_3$. This is consistent with what is found by \cite{Jeffries2006}, who already notice the presence of two kinematics components towards the cluster. 
The kinematic properties of group $B_3$ are similar to those of groups D,  $B_0$ (not located in the Belt region, see Fig. \ref{fig:fig4}),  and $B_5$. We notice that group $B_2$'s  velocity dispersion is  large ($\sim 1.6 \, \mathrm{km \, s^{-1}}$) compared for instance to that of group $B_3$ ($0.41 \pm 0.02 \, \mathrm{km \, s^{-1}}$).
The proper motion distribution shows indeed some substructures, which cause the large value of the velocity dispersion. As mentioned above, the presence of kinematic substructure may indicate the co-existence of groups with different kinematics in the same area. An inspection of group $B_2$'s 3D configuration (see Fig. \ref{fig:fig3}, in particular the $X-Y$ projection) shows that the source distribution is not uniform, and seems to be divided into (at least two) elongated structures.
\item  Group $B_5$ is located below the Belt, towards $\eta$ Ori, and shares similar kinematics with group $B_3$, although they seem to be well separated in space (see Fig. \ref{fig:fig3} and \ref{fig:fig4}).
The proper motion distribution shows two clumps, similar to what is observed towards 25 Ori. We separate the
the smaller clump, which we refer to as $B_{5,b}$ by using simple cuts in proper motion space:
\begin{align}\label{eq:eq12}
    0.3 \, \mathrm{mas \, yr^{-1}} & < \mu_{\alpha*} <2. \, \mathrm{mas \, yr^{-1}};  \nonumber\\ 
    -0.8 \, \mathrm{mas \, yr^{-1}}   & < \mu_{\delta}  < 0.3 \, \mathrm{mas \, yr^{-1}}.
\end{align}
In contrast to what we have done for group $B_{0,b}$, here we apply the conditions of Eq. \ref{eq:eq12} to  all the sources in the Belt region, and not just those within the $f(D) = 0.5$ level of the 3D density map. This is the reason why the number of sources is higher than for group E (see Table \ref{tab:table4}). This choice is motivated by the fact that the visual inspection of the proper motion  diagram suggests that the clump is more extended and the number of sources is larger than what found by DBSCAN.  Further, the number of sources of the smaller clump is too small to retrieve the kinematic parameters accurately. The parameters estimated by the kinematic modelling code (see Table \ref{tab:table4}) show that group $B_5$ and group $B_{5,b}$ have different kinematic properties, while having similar parallaxes.
Comparing Fig. \ref{fig:fig8} and Fig. \ref{fig:fig13} one can notice that the region defined in Eq. \ref{eq:eq12} also includes sources classified as members of group $B_2$. The sky distribution of sources belonging to group $B_{5,b}$ (see Fig. \ref{fig:fig16}) shows indeed some sources clustering around $\sigma$ Ori. Most of the sources however are located in the same region as group $B_{5}$, although they are spread throughout the entire longitudinal extent of the Belt region. This seems to suggest that group $B_{5,b}$ is more extended than the Belt region, especially to lower galactic latitudes and longitudes. Similar conclusions can be drawn after studying the 3D distribution of group $B_{5,b}$ (Fig. \ref{fig:fig16}): some sources clump in the same area as group $B_2$ and $B_4$ ($\sigma$ Ori), while others are located closer to group $B_5$. This explains why DBSCAN does not separate successfully groups $B_5$ and $B_{5,b}$: their members show different kinematics but are mixed in space.

\begin{figure}
    \centering
    \includegraphics[width =\hsize]{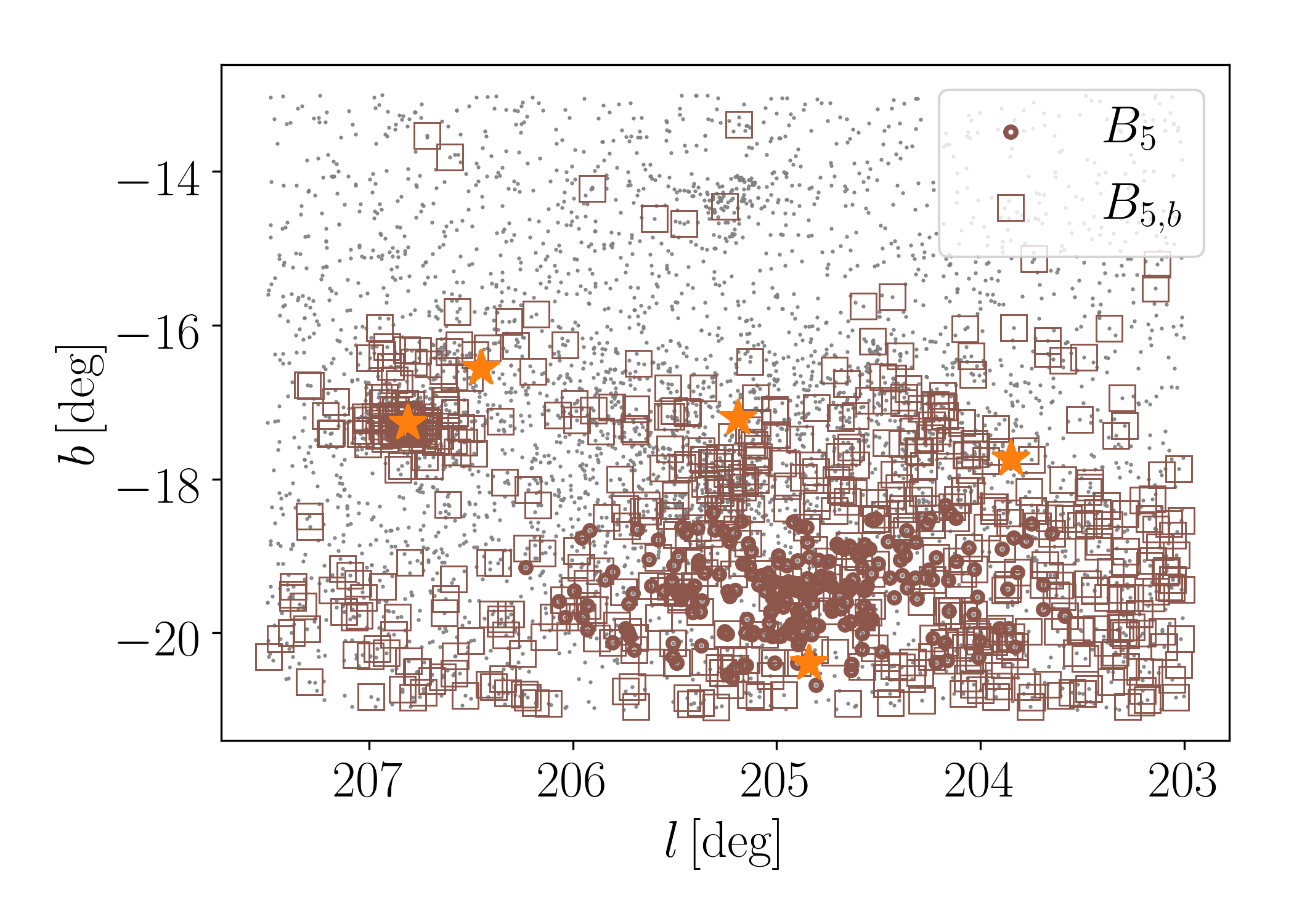}\\
    \includegraphics[width =\hsize]{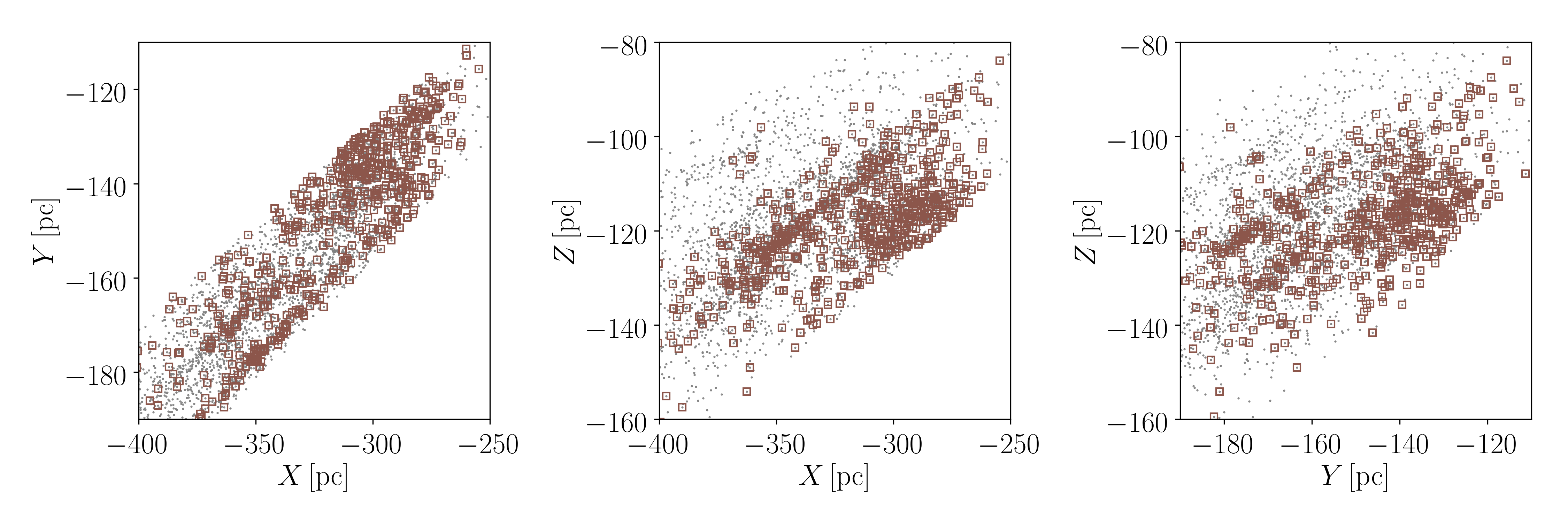}\\
   \caption{Top: Sky distribution of the sources belonging to group $B_{5,b}$ (brown empty squares), group $B_5$ (brown dots), and all the sources in the Belt region defined in the text (grey dots). The orange stars mark the position of $\sigma$ Ori, $\zeta$ Ori, $\epsilon$ Ori, $\eta$ Ori, and $\delta$ Ori.
   Bottom: distribution in 3D galactic coordinates of group $B_{5, b}$ (brown squares), and of all the sources belonging to the Belt region (grey dots).
    }
    \label{fig:fig16}
\end{figure} 

\item Group E is the most distant group in the entire Orion region (see Table \ref{tab:table4} and Fig. \ref{fig:fig3}). Since not many radial velocity measurements are available, the kinematic properties are determined with less accuracy than for the other groups, especially in the $y_I$ direction. While $v_{y,I}$  is comparable with those of  group A, C, $B_1$, $B_2$, $B_4$ (and $B_{7,b}$ and $B_8$, see below), the $v_{x,I}$ component is different from the other groups. 
As for group $B_5$, the proper motions seem to be divided in two clumps,  one of which 
does not correspond to any other DBSCAN groups. We select group $B_{8}$ by applying the following conditions: \begin{align}\label{eq:eq13}
    -2.2 \, \mathrm{mas \, yr^{-1}} & < \mu_{\alpha*} < -0.5\, \mathrm{mas \, yr^{-1}};   \nonumber\\
    0.4 \, \mathrm{mas \, yr^{-1}}  & < \mu_{\delta}  < 2.2 \, \mathrm{mas \, yr^{-1}}.
\end{align}
Similarly as for group $B_{5,b}$, and with the same motivations, we consider again all the sources in the Belt region. The estimated kinematic parameters are reported in Table \ref{tab:table4}. The source distribution in the sky and in 3D Cartesian space is shown in Fig. \ref{fig:fig12}, compared to that of group E. The sources are loosely distributed in the entire Belt region, although they seem to clump next to group E. 
\item DBSCAN identifies only 30 sources belonging to group $B_7$, none of them with a measured radial velocity, therefore the kinematic modelling code does not succeed in determining reliable parameters. 
Similarly to  what was found for group $B_{5,b}$ and $B_{8}$, when considering all the stars in the Belt area, we notice that many more sources clump in the same proper motion region that are excluded when we apply the condition $f(D) > 0.5$ or that are classified as 'noise' stars by DBSCAN.
We therefore select group $B_{7b}$  according to the following equations (see Fig. ):
\begin{align}\label{eq:eq14}
    -2.2 \, \mathrm{mas \, yr^{-1}} & < \mu_{\alpha*} <-0.5\, \mathrm{mas \, yr^{-1}};  \nonumber\\ 
    -2 \, \mathrm{mas \, yr^{-1}}   & < \mu_{\delta}  < 0.4 \, \mathrm{mas \, yr^{-1}}.
\end{align}
The number of sources is now much larger (see Table \ref{tab:table4}), and the parameters can be accurately determined. 
Fig. \ref{fig:fig13} shows the source distribution in the sky and in Cartesian galactic coordinates. We notice that the sources are distributed in the sky towards the reflection nebulae M78 and NGC 2071, where two groups of young stars are present and towards the centre of the Belt. 
\item Figure \ref{fig:fig10} shows the dust distribution towards the Belt region, where a bubble is visible \citep[see][]{Ochsendorf2014, Ochsendorf2015}. Some of the groups we identified might be responsible for the origin of the Belt bubble. In particular groups E and $B_8$ are located in the sky within the dust structure shown in Fig. \ref{fig:fig10}, at different distances. Group $B_8$ is slightly more diffuse than the bubble, but the central over-density is still located within the bubble boundaries. The stellar winds and the supernova explosions coming from these groups  might be responsible for the creation of the bubble itself. 

\begin{figure}
    \includegraphics[width =\hsize]{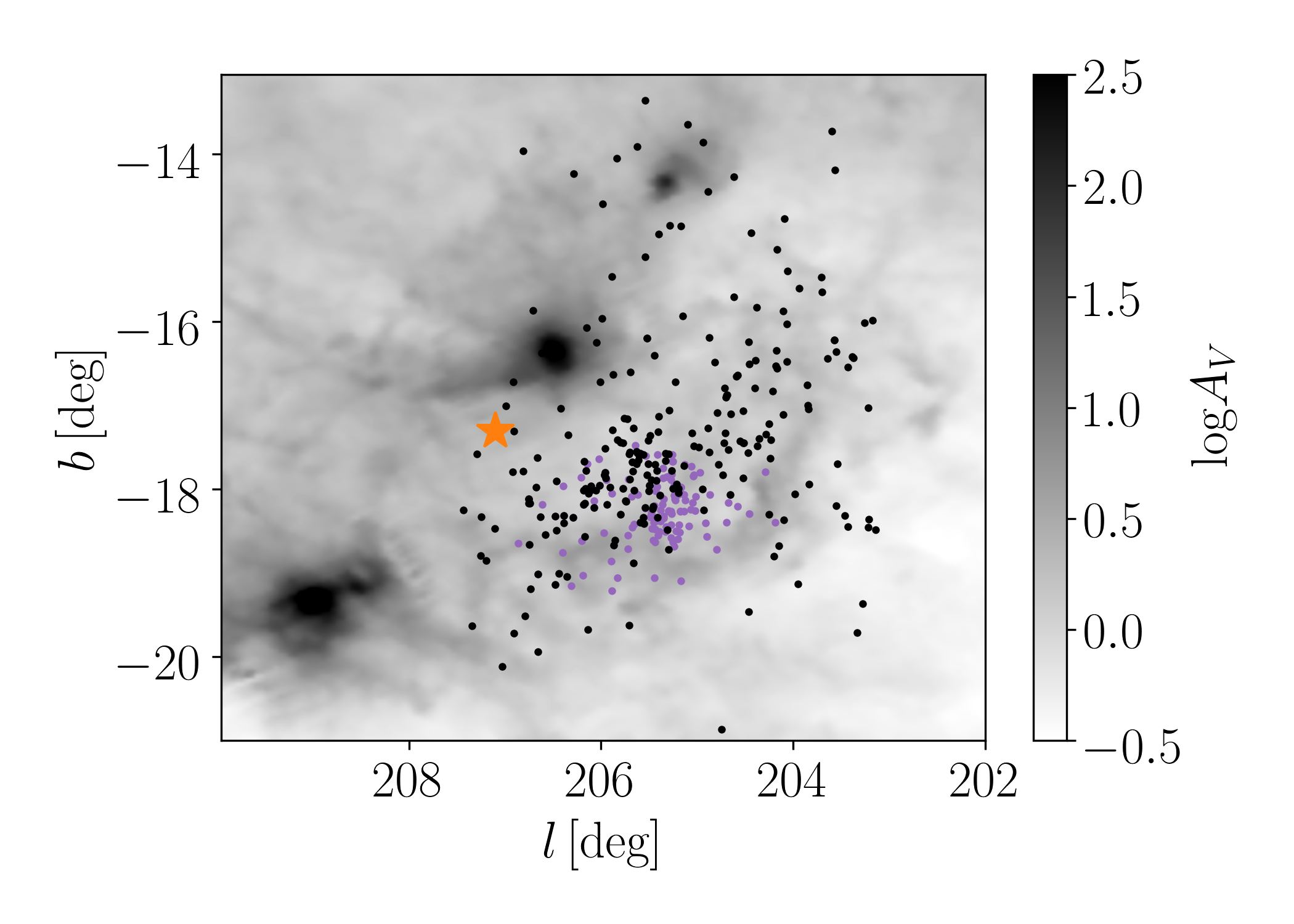}
    \caption{\textit{Planck} data and groups E (purple dots) and $B_8$ (black dots). The orange star represents $\sigma$ Ori.}
    \label{fig:fig10}
\end{figure}

\begin{figure}
    \centering
    \includegraphics[width =\hsize]{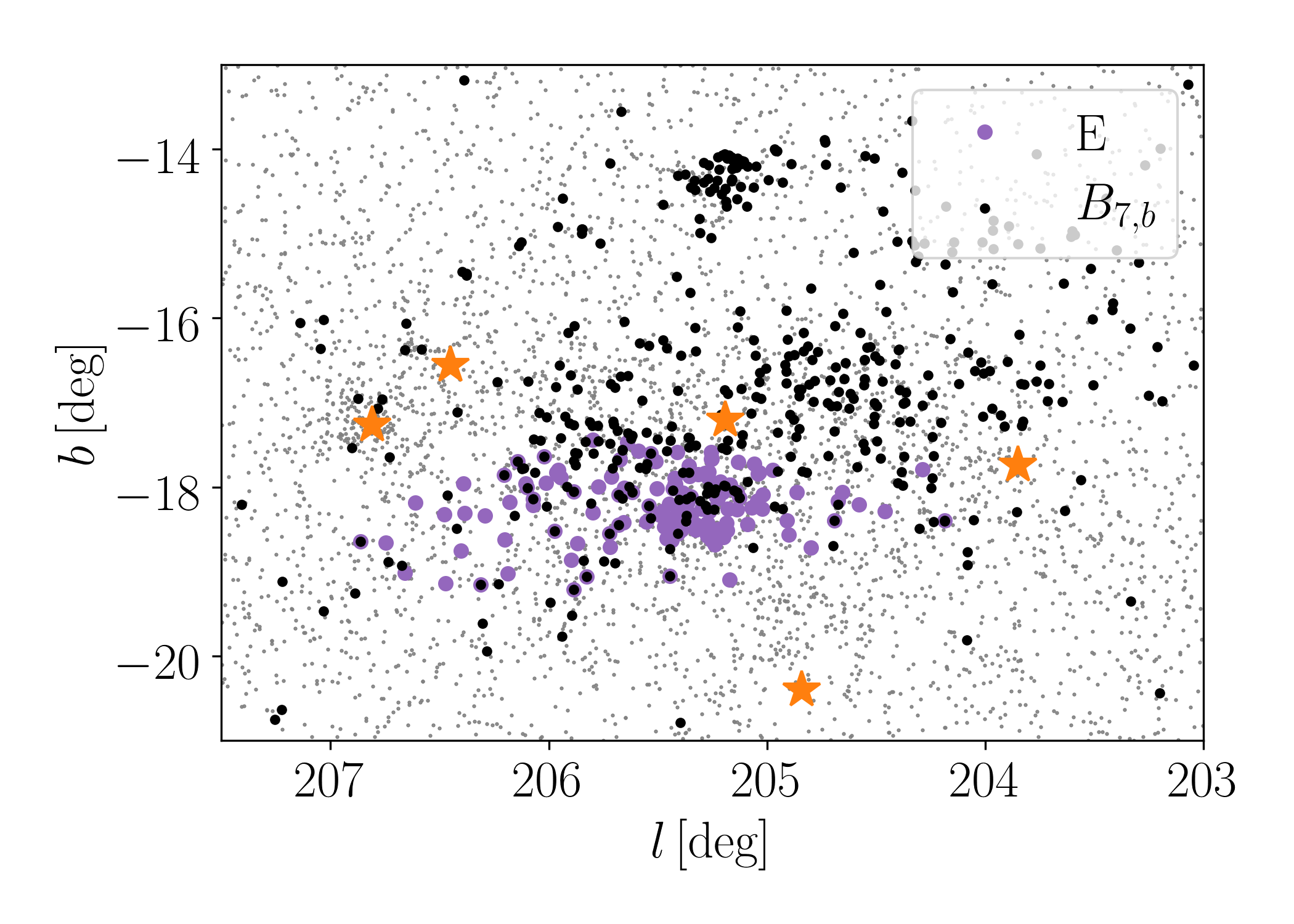}\\
    \includegraphics[width =\hsize]{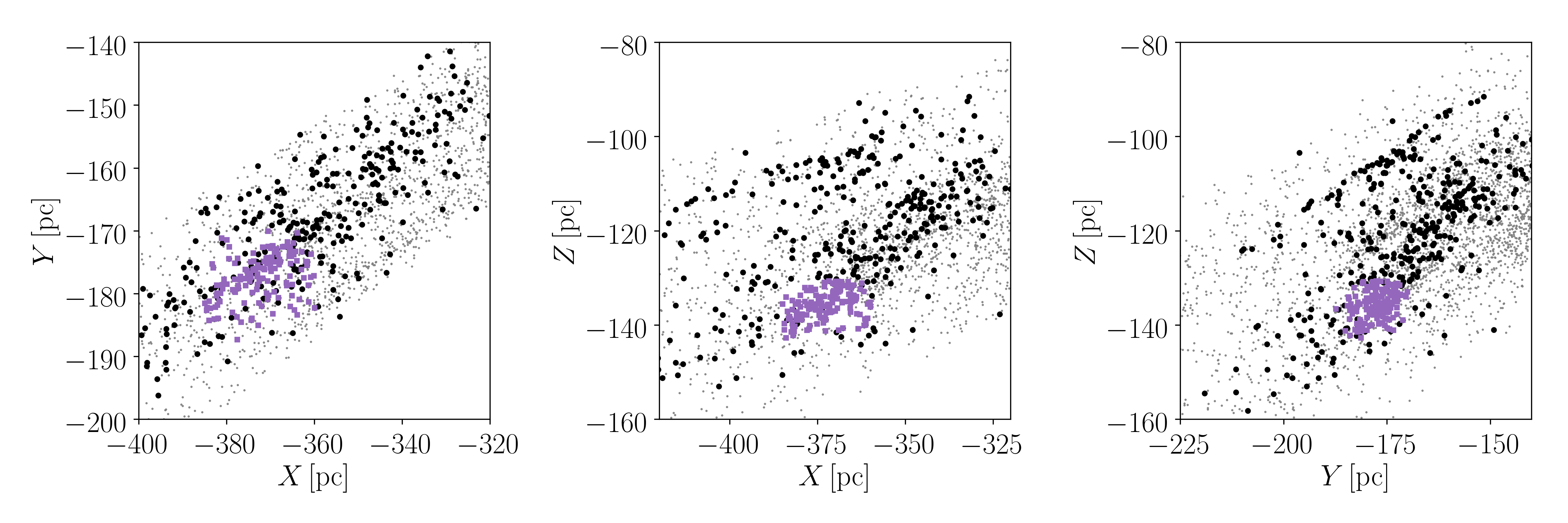}
    \caption{Distribution in the sky (top) and in 3D space (bottom) of the stars belonging to group $B_{7,b}$ (black dots) compared to those in group E (purple dots) The grey dots represent all the sources in the Belt region. The orange stars are the same as defined in Fig. \ref{fig:fig11}.
    }
    \label{fig:fig13}
\end{figure}
\begin{figure}
    \centering
    \includegraphics[width =\hsize]{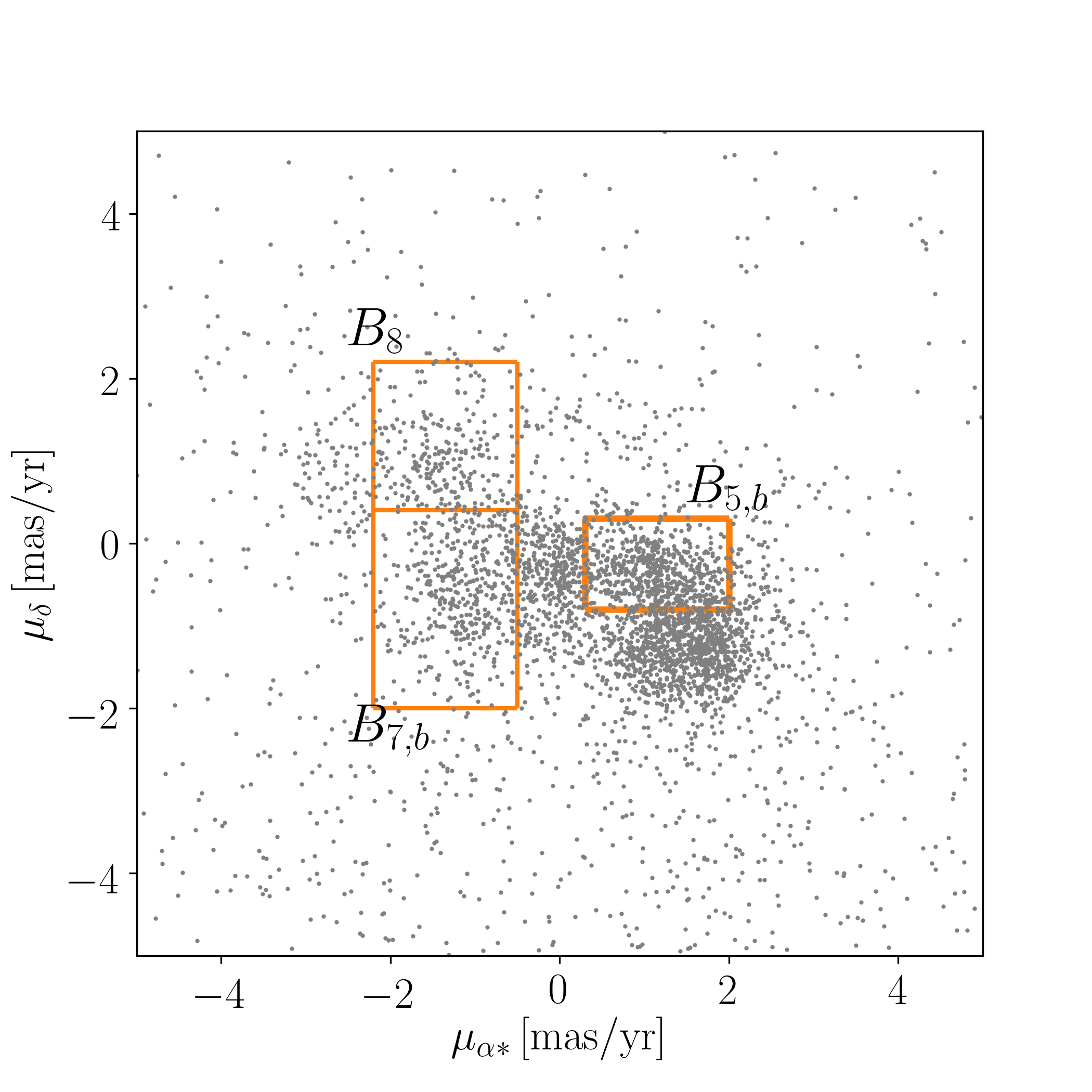}
    \caption{Proper motion diagram of all the sources in the Belt region. The orange rectangles are those defined in Eq. \ref{eq:eq12} and \ref{eq:eq13}.}
    \label{fig:fig13b}
\end{figure}
\begin{figure}
    \centering
    \includegraphics[width =\hsize]{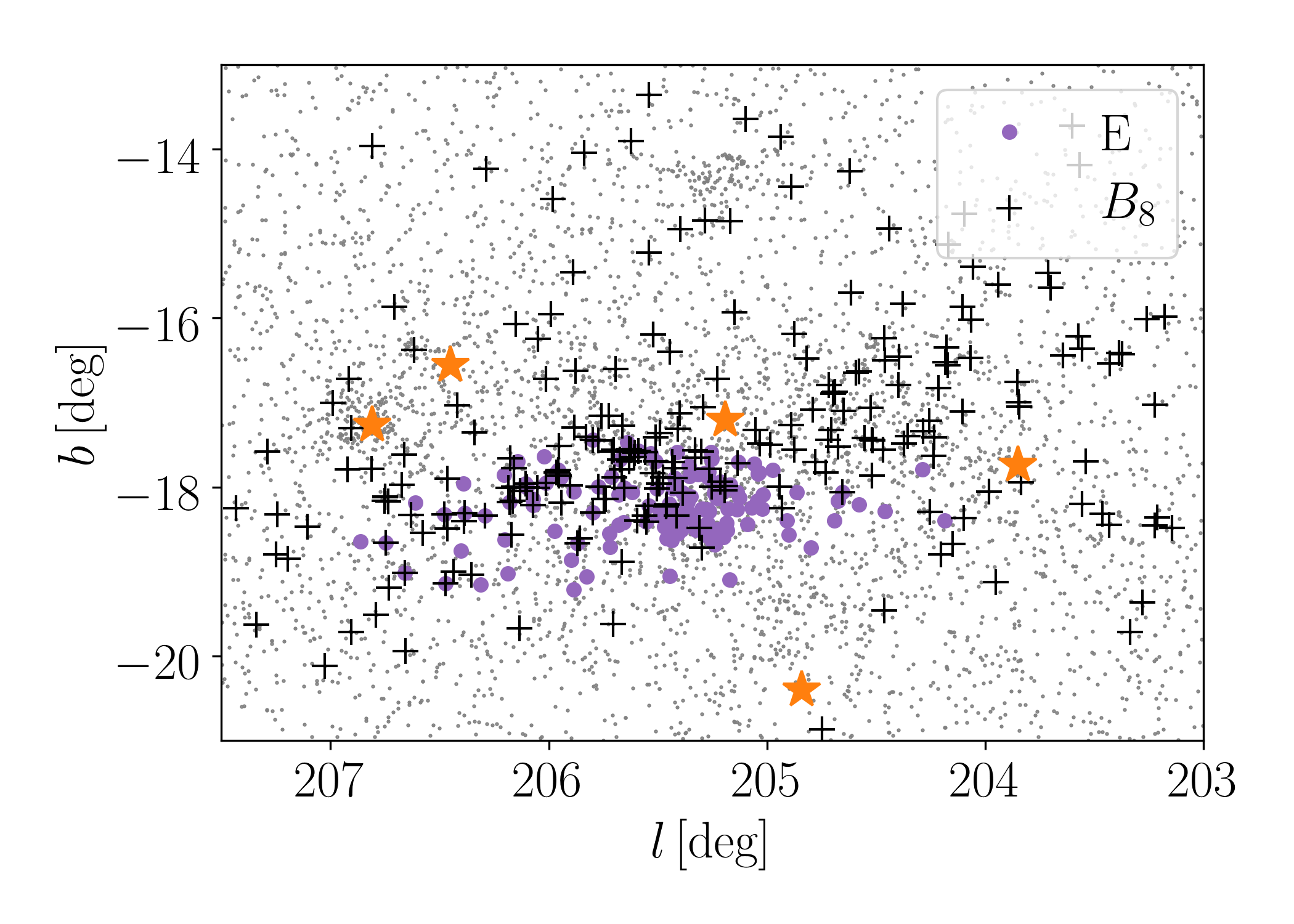}\\
    \includegraphics[width =\hsize]{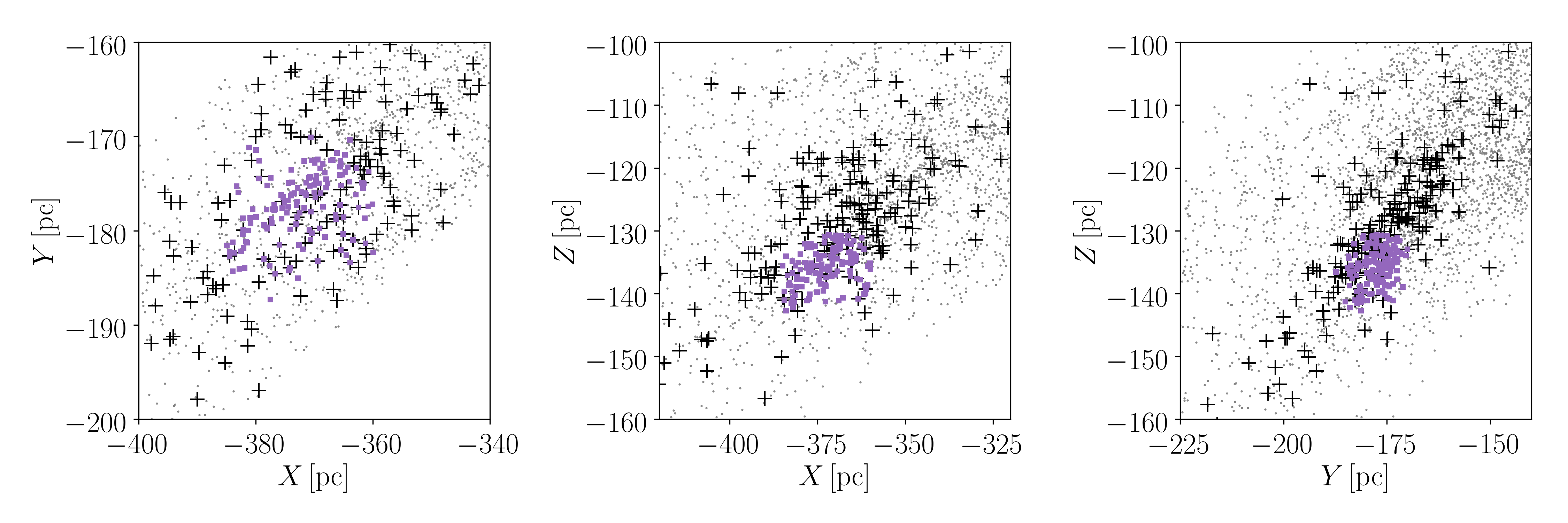}
    \caption{Distribution in the sky (top) and in 3D space (bottom) of the stars belonging to group $B_8$ (black crosses) compared to those in group E (purple dots). The orange stars are the same as defined in Fig. \ref{fig:fig11}.
    }
    \label{fig:fig12}
\end{figure}
\end{itemize}

\subsubsection{Orion A}
\begin{figure*}
    \centering
    \includegraphics[width =\hsize]{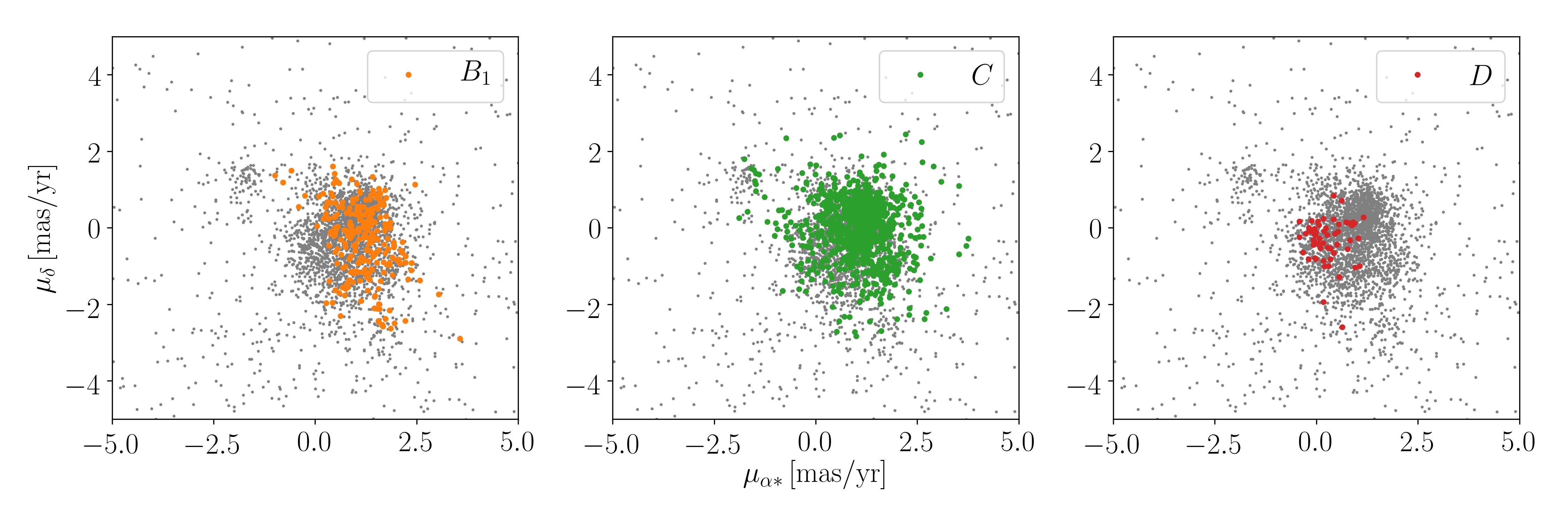}\\
    \caption{Proper motion diagram of the sources towards the Orion A molecular cloud, for group $B_1$ (orange dots in the left panel), group C (green dots in central panel), and group D (red dots in right panel). The grey dots represent the proper motions of all the sources in the Orion A region (see text).
    }
    \label{fig:fig14}
\end{figure*}

\begin{figure*}
    \centering
    \includegraphics[width =\hsize]{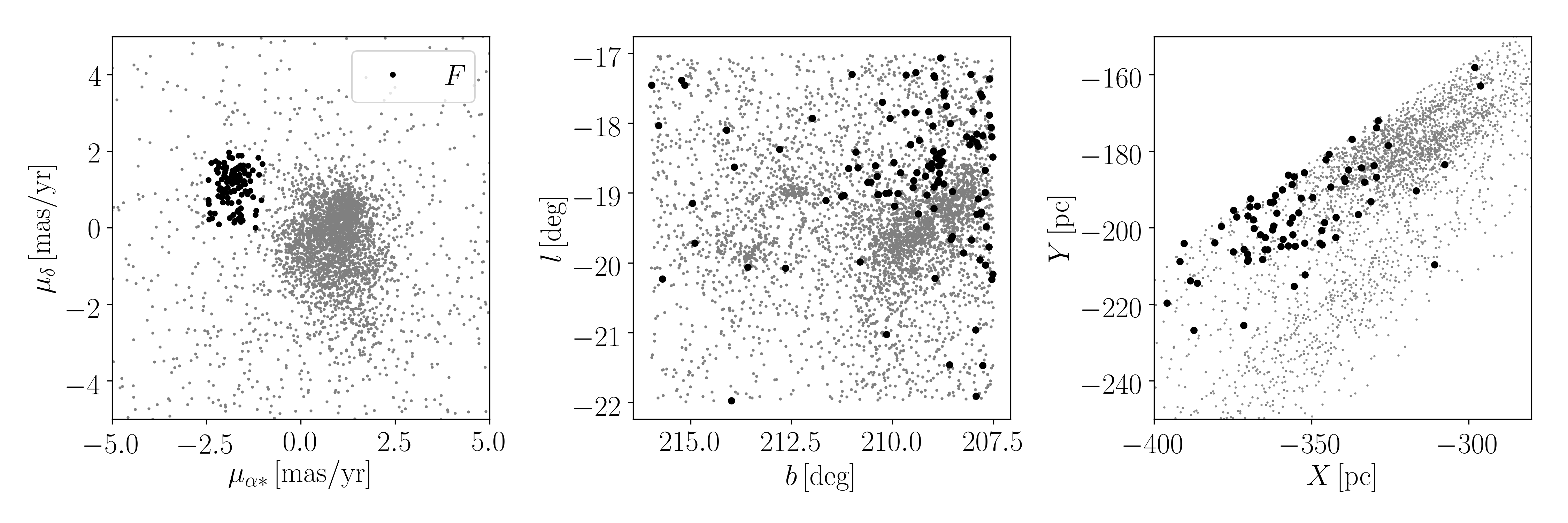}\\
    \caption{Proper motion (left), sky distribution (centre), and distribution in the $X-Y$ plane in Cartesian galactic coordinates of all the sources in the Orion A region (grey dots) and of those selected through Eq. \ref{eq:eq15} (black dots).
    }
    \label{fig:fig15}
\end{figure*}

The DBSCAN groups associated with the Orion A molecular cloud are those labelled $B_1$, C, and D. Group $B_1$ and C nearly occupy the same position in the sky and share very similar kinematic properties (see Table \ref{tab:table4}), however they are at different distances, with group $B_1$ being closer to the Sun than group C.
This poses interesting questions about their origin: the two groups might be identified separately by DBSCAN just because of a local under-density of sources. In this case, the Orion A cloud would be even more elongated along the line of sight than previously thought \citep{Grossschedl2018}. 
The radial velocities of the embedded sources in the Orion A molecular cloud are tightly related to the motion of the molecular gas in the cloud \citep{Hacar2016}. So, if the foreground is moving as the stars in the cloud, and stars in the cloud are coupled to the gas, the foreground group might have  originated from the same cloud complex. 
The proper motion diagram of the three groups is shown in Fig. \ref{fig:fig14}. We define the Orion A region as:
\begin{align}
    207.5^{\circ}  < l < 216^{\circ},  \qquad 
    -22^{\circ}  < b < -17^{\circ}.
\end{align}
The proper motions of all the sources (grey dots in Fig. \ref{fig:fig14}) in the region  show a clump in $\mu_{\alpha*}, \mu_{\delta} \sim (-2., 1)$ (see also left panel of Fig. \ref{fig:fig15}). We select the sources with proper motions:
\begin{align}\label{eq:eq15}
    -2.5 \, \mathrm{mas \, yr^{-1}} & < \mu_{\alpha*} <-1.\, \mathrm{mas \, yr^{-1}};  \nonumber\\ 
    0 \, \mathrm{mas \, yr^{-1}}   & < \mu_{\delta}  < 2. \, \mathrm{mas \, yr^{-1}},
\end{align}
(black dots in Fig. \ref{fig:fig15}) and we study their distribution in the sky and on the $X-Y$ plane in galactic Cartesian coordinates. We label this group as group F.  Fig. \ref{fig:fig15} (centre) shows that the sources are loosely distributed in the Orion A region, and seem to cluster at $(l, b) \sim (209, -19)$.  Fig. \ref{fig:fig15} (right) show that the members of group F are loosely spread at larger distances than the sources associated with the Orion A molecular cloud. We run the kinematic modelling on group F and we find the parameters reported in Table \ref{tab:table4}.
We compare the proper motions of group F with those of the other groups, and we notice that they are roughly the same as those of group $B_8$ (see Fig. \ref{fig:fig13}). Nevertheless the results of the kinematic modelling for the two groups are quite dissimilar. This could be due to the fact that, for both groups, the number of stars with measured radial velocity is small, and therefore the 3D velocity is not well constrained.
An inspection of the parallax distribution of group F also shows a number of sources with small parallax ($\varpi < 1.9 \, \mathrm{mas}$), which are most likely field contaminants.  


\begin{table*}
    \renewcommand{\arraystretch}{1.5}
    \caption{Results of the kinematic modelling for the groups identified by DBSCAN.
    The first column indicates the region in the sky where the groups are located.
    }
    \label{tab:table4}
    \begin{center}

    \begin{tabular}{llllllllll} 
        & \# & $N$ & $N_{RV}$ & $v_{x, I} \, [\mathrm{km \, s^{-1}}]$ & $v_{y, I} \, [\mathrm{km \, s^{-1}}]$ & $v_{z, I} \, [\mathrm{km \, s^{-1}}]$ & $\sigma_v \, [\mathrm{km \, s^{-1}}]$ & $\kappa \,   [\mathrm{km \, s^{-1} \, pc^{-1}}]$ & $\varpi \, [\mathrm{mas}]$  \\
      \hline
      \hline
      $\lambda$ Ori & A &  296 & 81 & 0.75 $\pm $ 0.05 &  27.4 $\pm $ 0.1 &  0.5 $\pm $ 0.05 &   0.73 $\pm $ 0.02 &  0.122 $\pm $ 0.007  &  $2.48^{0.07}_{0.06}$\\ 
      
      Orion A & C & 1059 & 489 &  0.94 $\pm $ 0.06 & 27.1  $\pm $ 0.09 & -2.68 $\pm $ 0.06 &  1.63 $\pm $ 0.03 & 0.07 $\pm $ 0.006  & $2.5^{0.06}_{0.08}$\\
      
      Orion A & D & 69 & 50 & 1.1 $\pm $ 0.1 &  21.8 $\pm $ 0.2 & -3.7 $\pm $ 0.1 &  0.98 $\pm $ 0.07 & -0.02 $\pm $ 0.02 &  $2.37^{0.02}_{0.03}$\\

      Belt & E & 150 & 10 &  4.6 $\pm $ 0.3 & 32.4 $\pm $ 2.7 & -1 $\pm $ 0.1 &  1.28 $\pm $ 0.06 & -   & $2.31^{0.03}_{0.04}$ \\

      25 Ori & $B_{0}$ & 710 & 73  &  0.41 $\pm $ 0.04  & 19.2 $\pm $ 0.2 & 0.045 $\pm $ 0.03 &  0.74 $\pm $ 0.02   & - & $2.84^{0.08}_{0.08}$\\

	  25 Ori & $B_{0b}$ & 54 & 4  &  4.8 $\pm $ 0.4  & 25. $\pm $ 2.8 & 2.1 $\pm $ 0.1 &  0.38 $\pm $ 0.04   & - & $2.78^{0.06}_{0.06}$  \\

	  Orion A & $B_1$ & 265 & 73 &  0.8 $\pm $ 0.1 & 26.8 $\pm $ 0.2 & -3.0 $\pm $ 0.1 &  1.55 $\pm $ 0.05 & 0.03 $\pm$ 0.02   & $2.74^{0.04}_{0.07}$ \\

      Belt & $B_2$  & 174 & 48 &  0.17 $\pm $ 0.14 &  27.84 $\pm $ 0.3 &  -2.6 $\pm $ 0.1 &  1.6 $\pm $ 0.07 & 0.08 $\pm $ 0.03  & $2.48^{0.04}_{0.05} $\\

      Belt & $B_3$ & 290 & 48 &  -0.74 $\pm $ 0.03  & 22.5 $\pm $ 0.2 &  -2.69 $\pm $ 0.05 &  0.79 $\pm $ 0.03 & -  &  $2.78^{0.07}_{0.08}$\\

      Belt & $B_4$ & 46 & 10 & -0.2 $\pm $ 0.3  & 26.5 $\pm $ 0.6 &  -3.1 $\pm $ 0.2 &  1.6 $\pm $ 0.1 & 0.07 $\pm $ 0.06 &  $2.62^{0.02}_{0.02}$\\
      
      Belt & $B_5$ & 248 & 12 &  0.9 $\pm $ 0.1  & 20.7 $\pm $ 0.9 &  -2.44 $\pm $ 0.06 &  0.7 $\pm $ 0.03 & -  &  $2.79^{0.09}_{0.06}$\\
      
      Belt & $B_{5,b}$ & 622 & 48 &  1.2 $\pm $ 0.1  & 24.8 $\pm $ 0.2 &  -1.42 $\pm $ 0.04 &  0.82 $\pm $ 0.02 & -  &  $2.75^{0.22}_{0.36}$\\
      
      25 Ori & $B_6$ & 40 & 0 &  - & -  &  - & - & - & - \\

      Belt & $B_7$ & 30 & 0 &  -  & - &  - & - & - & - \\

      Belt & $B_{7,b}$ & 441 & 63 &  4.9 $\pm$ 0.05  & 26.7 $\pm $ 0.2 & -1.55 $\pm$ 0.05 & 0.95 $\pm $ 0.03 & 0.024 $\pm$ 0.003   &  $2.36^{0.2}_{0.23}$\\

      Belt & $B_8$ & 245 & 18 & 5.8 $\pm$ 0.08 & 28.5 $\pm$ 0.4 & 1.5 $\pm$ 0.06 & 0.9 $\pm$ 0.03 & 0.05 $\pm$ 0.05 &  $2.34^{0.2}_{0.18} $    \\ 
	  
	  Orion A & F  & 116 & 17 & 5.8 $\pm$ 0.1 & 21.2 $\pm$ 0.5 & 0.3 $\pm$ 0.12 & 1. $\pm$ 0.06 & - &  $2.27^{0.2}_{0.48}$  

    \end{tabular}
     \end{center}
\end{table*}

\begin{table}
   \renewcommand{\arraystretch}{1.5}
    \caption{Age estimates for the groups identified in Section 3 and 4. The column $\mathrm{\log(age/yr)}$ ($\tau$) indicates the (log-)age estimated by the isochrone fitting procedure. The column $\tau_{exp}$ indicates the expansion ages determined by using the formula $\tau_{exp} = 1/(\gamma \, \kappa)$ for the groups for which it is possible to determine the expansion parameter $\kappa$. The number of stars $N$ is different than in Table \ref{tab:table1} because by applying the kinematic modelling we remove kinematic outliers from the groups.
    }
    \label{tab:table5}
    \begin{tabular}{llllll} 
       \# & $N$ & $\mathrm{\log(age/yr)}$ & $\tau \, [\mathrm{Myr}]$ & $A_V \, [\mathrm{mag}]$ & $\tau_{exp} \, \mathrm{[Myr]}$ \\
      \hline
      \hline
      A &  274 & $6.75^{0.03}_{0.01}$ & $5.6^{0.4}_{0.1}$ & 0.4  & 8.0  \\ 
      
      C & 943 & $6.9^{0.03}_{0.01}$ & $8^{0.5}_{0.04}$ & 0.2  &  14.0 \\
      
      D & 60 &  $6.85^{0.03}_{0.02}$ & $7^{0.6}_{0.2}$ &  1.3  &  - \\

	  E & 139 &  $7.05^{0.04}_{0.005}$ & $11.2^{1}_{0.1}$ &  0.5 & - \\

	  $B_{0}$ & 622 & $7.05^{0.04}_{0.005}$ & $11.2^{1}_{0.1}$  & 0.2 &  -  \\
	  
	  $B_{0b}$ & 44  & $7.15^{0.1}_{0.004}$ & $14^{3}_{0.25}$  & 0.4   & -  \\

	  $B_1$ & 246 & $7.0^{0.03}_{0.01}$ & $10^{0.7}_{0.23}$ &  0.4 &  32.6\\

      $B_2$  & 154 & $6.6^{0.03}_{0.01}$ & $4^{0.3}_{0.1}$ & 0.3 & 12.2  \\
      
      $B_3$ & 221 & $6.9^{0.04}_{0.01}$ & $8^{0.7}_{0.04}$ & 0.2  & -\\
      
      $B_4$ & 44 & $6.6^{0.03}_{0.01}$ & $4^{0.3}_{0.1}$ & 0  & 14  \\
      
      $B_5$ & 234  & $6.9^{0.04}_{0.01}$ & $8^{0.7}_{0.04}$ & 0.2  & -  \\
      $B_{5,b}$  & 605 &  $7.05_{0.005}^{0.03}$ & $11.2^{1}_{0.1}$ & 0.2 & - \\
      
      $B_{7,b}$ & 418 & $7.05^{0.04}_{0.005}$ & $11.2^{1}_{0.1}$ & 0.3  & 40 \\

	  $B_8$ & 237 &  $7.15^{0.04}_{0.004}$ &  $14^{1.5}_{0.25}$ & 0.3 & - \\
	  
	  F  & 108 &   $7.05^{0.03}_{0.005}$ & $11.2^{1}_{0.1}$ &  0.3 & -
 
    \end{tabular}
\end{table}

\section{Ages}\label{sec:5}
We determine ages ($\tau$) and extinctions ($A_V$) of the groups we identified by performing an isochrone fit based on a maximum likelihood approach similar to the methods described in \citet{Jorgensen2005},  \citet{Valls-Gabaud2014}, and \citet{Zari2017}.
\newline
Assuming independent Gaussian errors on all the observed quantities we can write the likelihood for a single star to come from an isochrone with certain properties $\bm{\theta} = (\tau, A_V, Z, ...)$, as:
\begin{equation}
L(\bm{\theta}, m) = \prod_{i=1}^n \left(\frac{1}{(2\pi)^{1/2}\sigma_i}\right) \times \exp{ \left(- \chi^2/2 \right)},
\end{equation}
with: 
\begin{equation}
\chi^2 = \sum_{i = 1}^n \left(\frac{q_i^{\mathrm{obs}}-q_i(\bm{\theta},m)}{\sigma_i}\right)^2,
\end{equation}
where $m$ is the stellar mass, $n$ is the number of observed quantities, and $\bm{q}^{\mathrm{obs}}$ and $\bm{q}(\bm{\theta},m)$ are the vectors of observed and modelled quantities. 
To take into account the fact that stars are not distributed uniformly along the isochrone, we weight the $j$th likelihood with a factor $w$ defined as:
\begin{equation}
    w = \sqrt{\frac{n_{redder \, j}}{n_{bluer \,j} + 1}}, 
\end{equation}
where $n_{redder}$ is the number of stars with $G_{BP}-G_{\mathrm{RP}}$ colour larger than that of the $j$th star and $n_{bluer}$ is the number of stars with $G_{BP}-G_{\mathrm{RP}}$ smaller than that of the $j$th star.
This choice gives larger weights to blue, massive stars,
to take into account that they are fewer than the low-mass members of the clusters.
\newline
The likelihood for $N$ coeval stars is just defined as:
\begin{equation}
    L_{combined}(\bm{\theta}, m) = \prod_{j=1}^N L_j(\bm{\theta}, m)^{w_j}
\end{equation}
Since we are interested in determining the ages and the extinctions of the groups, we fix  the metallicity to $Z = Z_{\odot} = 0.0158$ and we integrate Eq. 13 on the mass, so that the probability density function as a function of age $\tau$ and extinction $A_V$ is given by:
\begin{equation}
    L_{combined}(\tau, A_V) =  \prod_{j= 1}^N \int L_j(\tau, A_V, m)dm 
\end{equation}

To perform the fit we compare the observed $G$ magnitude and $G_{BP}-G_{\mathrm{RP}}$ color to those predicted by the PARSEC \citep[PAdova and TRieste Stellar Evolution Code][]{Bressan2012, Chen2014, Tang2014}  library of stellar evolutionary tracks, using the passbands by \cite{Maiz2018}. We used isochronal tracks from $\mathrm{\log(age/yr) = 6.0}$ (1 Myr) to $\mathrm{\log(age/yr) = 8.0}$ (100 Myr), with a step of $\mathrm{\log(age/yr) = 0.05}$ , and from $A_V = 0 \, \mathrm{mag}$ to  $A_V = 2.5 \, \mathrm{mag}$ with a step of $0.1 \, \mathrm{mag}$.  
\newline
Our fitting procedure does not take into account the presence of unresolved binaries, the photometric variability of young stars, the presence of circumstellar material, or potential age spreads within single groups. These
effects can bias our age estimates and this issue is further discussed in Section 6.2

\subsection{Results}
\begin{figure}
    \centering
    \includegraphics[width =\hsize]{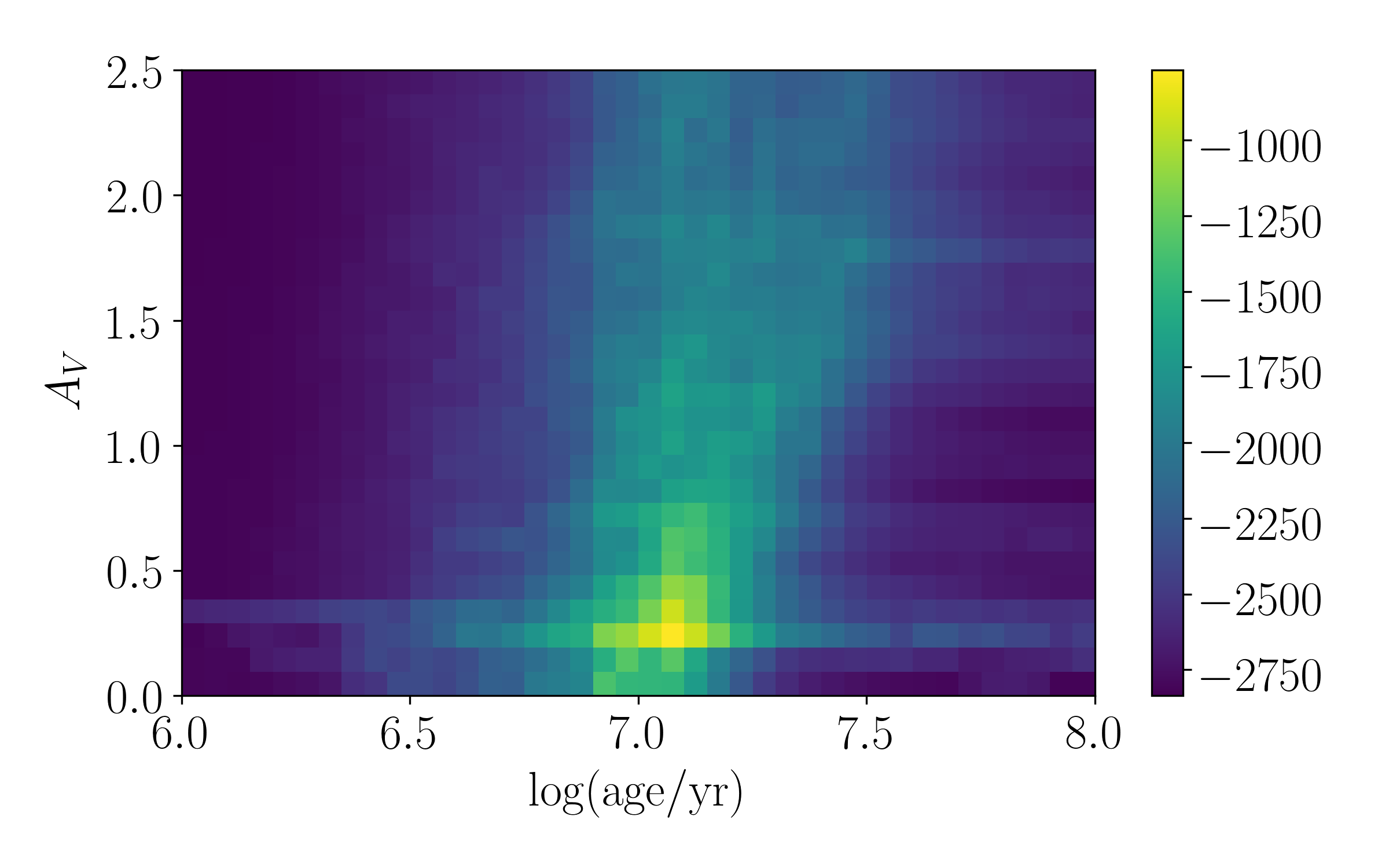}
    \caption{$\log L$ for the cluster $B_0$. Note the correlation between age and extinction.}
    \label{fig:fig5}
\end{figure}

\begin{figure}
    \centering
    \includegraphics[width =\hsize]{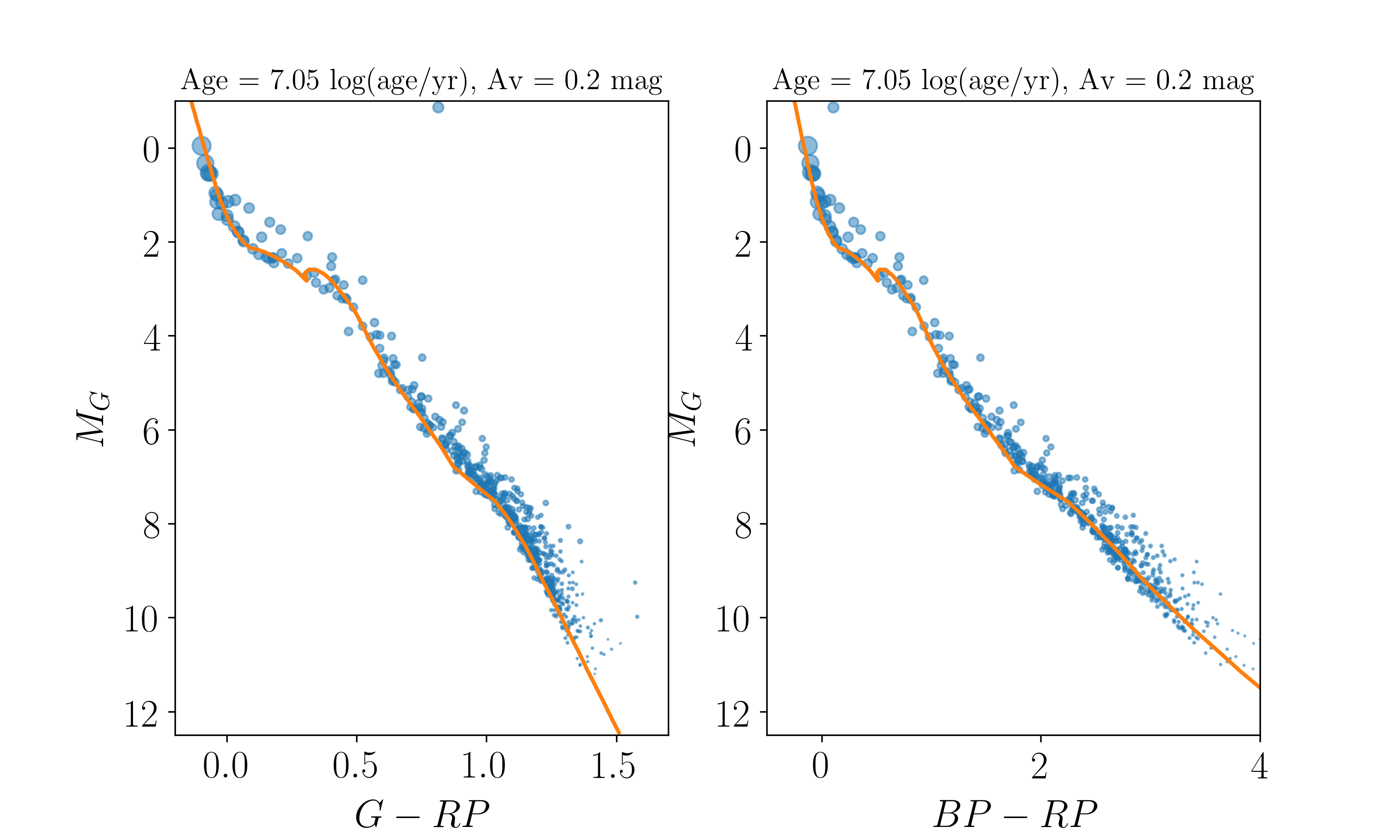}
    \caption{$M_G$ vs. $G - G_{\mathrm{RP}}$ (left) and $M_G$ vs. $G_{BP} - G_{\mathrm{RP}}$ (right) the colour magnitudes for group $B_0$. The symbol sizes represent the weights assigned to each star. The solid orange line represents the best fit isochrone.}
    \label{fig:fig6}
\end{figure}

We compute the age $\tau$ and the $A_V$ for the groups identified by DBSCAN, and for the groups we selected in
Section 4. The results are reported in Table \ref{tab:table5}.
Figures \ref{fig:fig5} and \ref{fig:fig6}  show the log-likelihood $\log L = \log L_{combined}(\tau, A_V)$ we obtain for group $B_0$, and the $M_G$ vs. $G - G_{\mathrm{RP}}$ (left) and $M_G$ vs. $G_{BP} - G_{\mathrm{RP}}$ (right) colour-magnitude diagrams (the colour-magnitude diagrams for the other groups are shown in Appendix B).
The orange solid line corresponds to the best-fitting isochrone. As mentioned above, we perform the fit using the $G_{BP} - G_{\mathrm{RP}}$ colour, and we show the colour-magnitude diagram in $G - G_{\mathrm{RP}}$ as a quality check. We adopt the maximum of $L_{combined}(\tau, A_V)$ as our best estimate of the stellar age, and we compute the confidence intervals by evaluating the 16th and the 84th percentiles after marginalizing over $A_V$. 
Figure \ref{fig:fig5} shows a correlation between age and extinction: at large extinction values the isochrones move towards redder colours, and soon they do not intersect the upper main sequence. However they still can fit the low pre-main sequence. 

\section{Dicussion}\label{sec:6}
In this section we summarise and comment the results obtained in the previous Sections and we put them in the broader context of the models of sequential star formation and triggering.
\noindent
\subsection{Kinematics}
By considering the $v_{y,I}$ velocities, we notice that we can roughly divide them in two groups, the first one with $v_{y,I} \sim 20 \, \mathrm{km \, s^{-1}}$  and the second one with $v_{y,I} \sim 26-27 \, \mathrm{km \, s^{-1}}$.
We observe a loose correlation between velocity and distances (the farthest objects are also the fastest), while there is no correlation between velocity and age or distance and age. 

In the kinematic modelling code we included isotropic expansion, however expansion could be an-isotropic, as observed for example by \cite{Cantat-Gaudin2018} and \cite{Wright2018}, although expansion due to residual gas expulsion is usually thought to be isotropic.
The expansion ages determined by using the formula $\tau_{\mathrm{exp}} = 1/(\gamma \kappa)$ give a loose indication  of the group ages, and confirm the age ordering obtained by the isochrone fitting procedure. The results of the simulations that we performed to test the kinematic modelling code (see Appendix A) showed that the expansion parameter $\kappa$ always resulted to be under-estimated, thus providing over-estimated expansion ages. This is consistent with the expansion ages obtained for the DBSCAN groups.

As mentioned in Section 3, by using the DBSCAN algorithm we preferentially select clusters that are dense in 3D space, and tend to neglect more diffuse groups. This effect is mitigated by the visual inspection of the proper motion diagrams of the DBSCAN groups, which we use to select groups with common kinematic properties that DBSCAN fails to retrieve. 
Further, one of the goals of the kinematic modelling code is to exclude outliers from the DBSCAN groups. Outliers are stars that do not share the same kinematic properties as the other cluster members: this implies that also stars that should be considered cluster members, such as binaries, are excluded from the DBSCAN groups. 
\newline
These considerations suggest that the groups that we analyse are not complete in terms of membership. The aim of this study is however to characterise the global properties of the stellar population in the Orion region. A more detailed analysis of the physical properties for which a complete membership list is important, such as the initial mass function,
is left to future studies.

\begin{figure*}
    \centering
    \includegraphics[width =0.33\hsize]{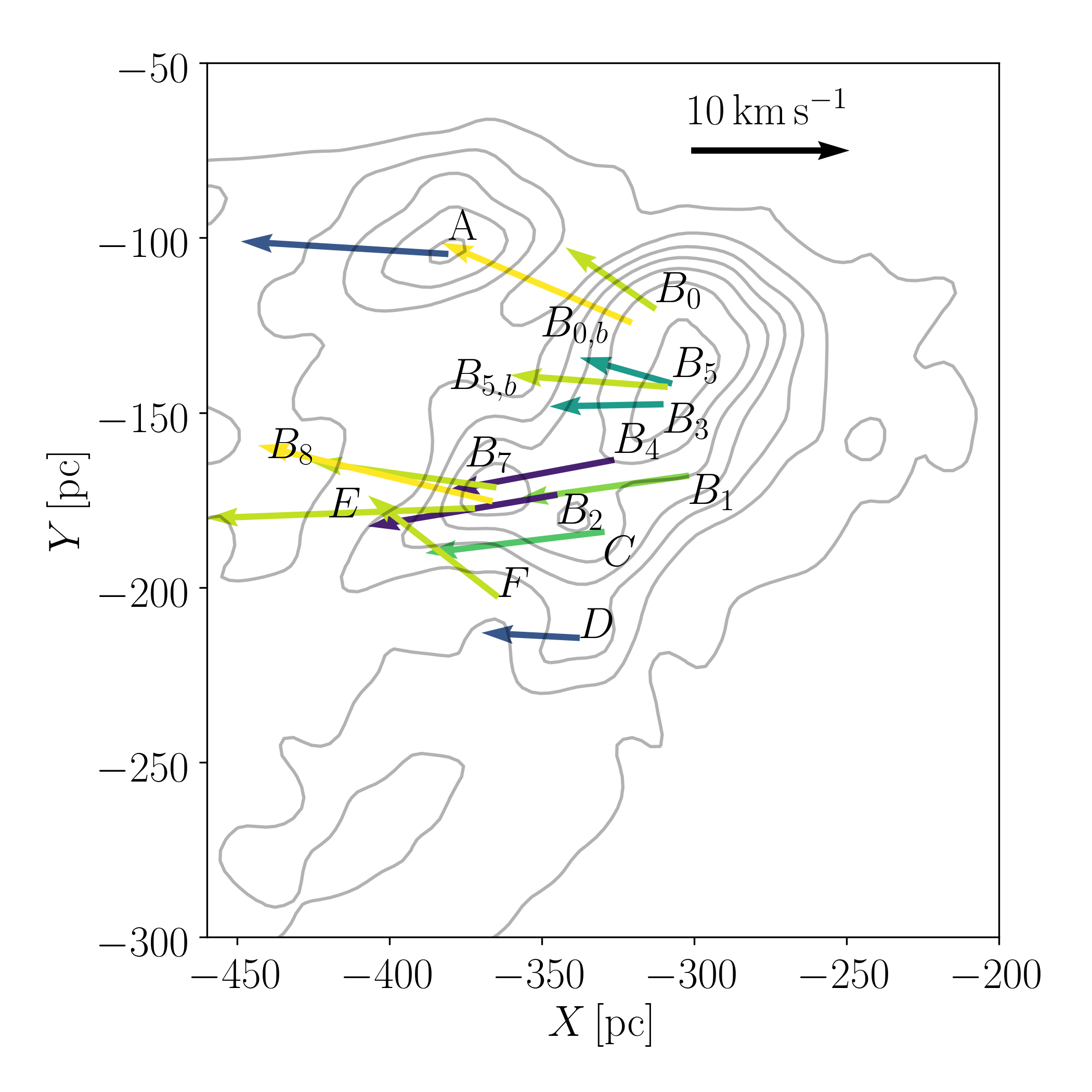}
    \includegraphics[width =0.33\hsize]{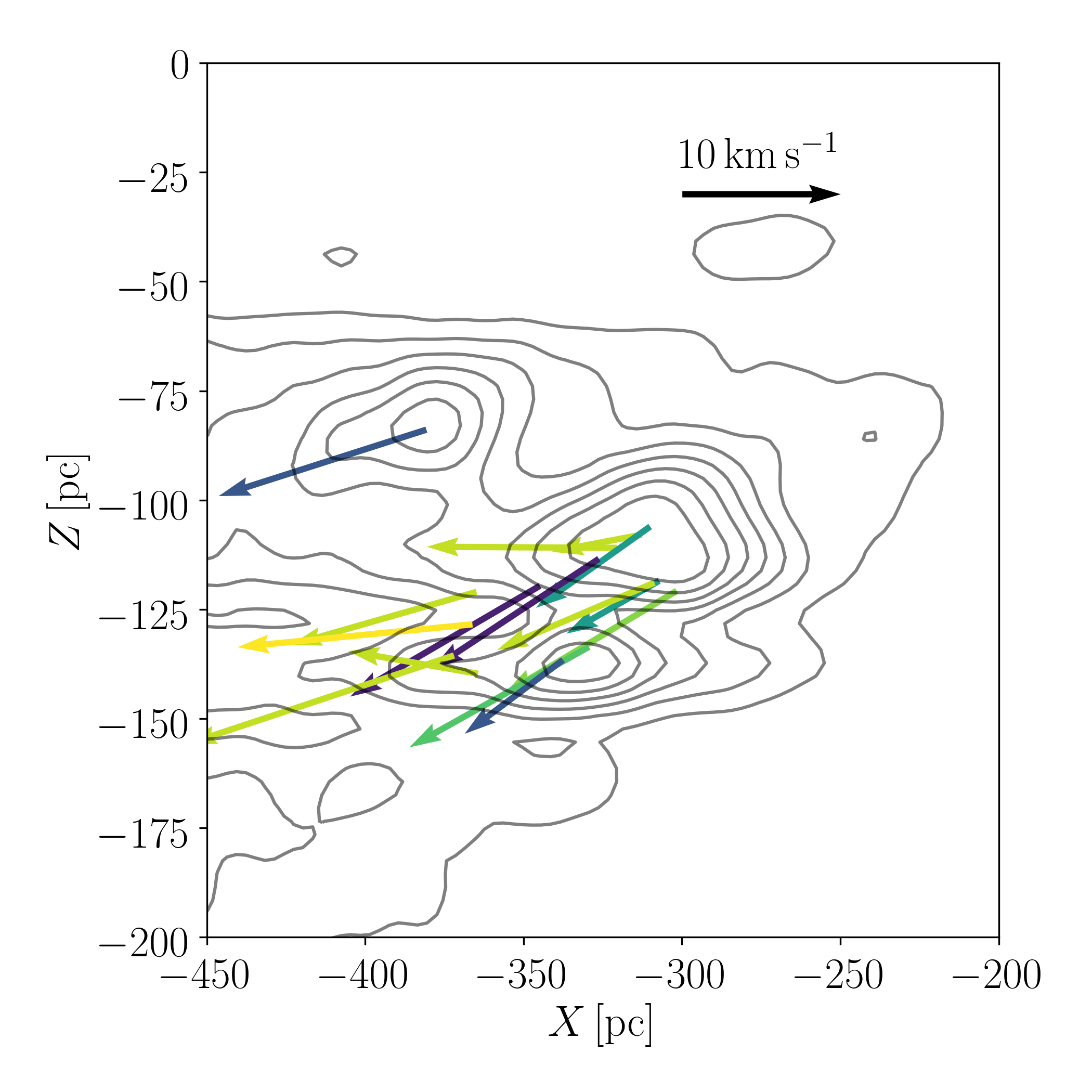}
    \includegraphics[width =0.33\hsize]{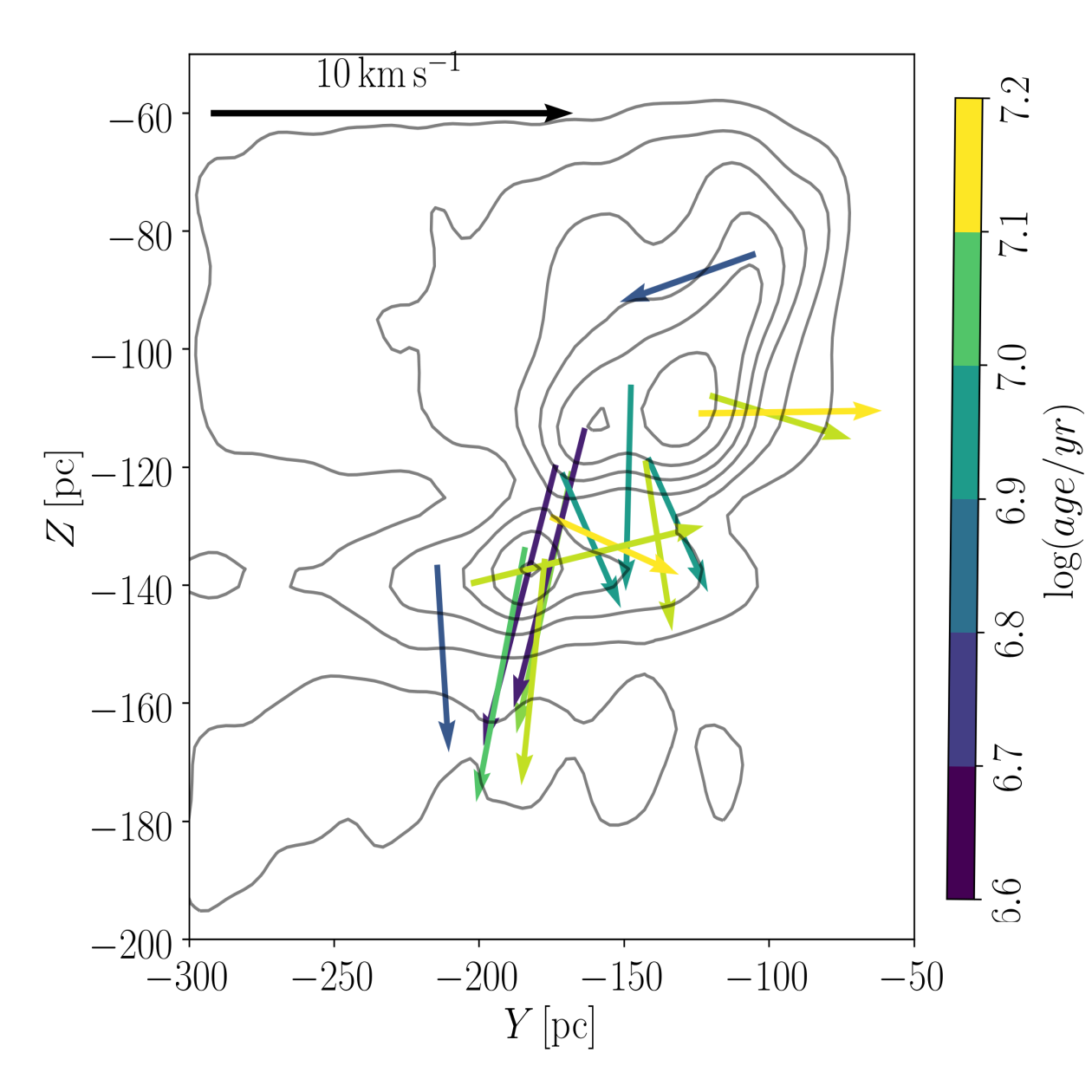}
    \caption{The contours represent the density distribution of the selected sources projected in the Galactic plane (left), in the $(X, Z)$ plane (centre), and in the $(Y, Z)$ plane (the Sun is at $(X, Y, Z) = 0,0,0$). The arrows represent the velocities (in Galactic coordinates) estimated in Section 3 for all the groups, and are corrected for the solar motion. The colours of the arrows represent the $\log(age/yr)$  obtained by fitting the colour-magnitude diagrams of the clusters in Section 4. }
    \label{fig:fig7}
\end{figure*}

\subsection{Ages}
The results obtained by fitting isochrones to the colour-magnitude diagrams of the groups isolated in Section 4 confirm the existence of the old population towards the 25 Ori group found by \cite{Kos2018}, which corresponds to our group $B_{0,b}$. \cite{Kos2018} derive an age of 20 Myr, while we obtain an age of 15 Myr. This could depend on the different extinction values used or by a slightly different membership list.
We also found that, towards the Belt, group E, $B_{5,b}$, $B_{7, b}$, and $B_{8}$ are older than 10 Myr, and that some older sources are also found in the Orion A region (group F). 
The population in front of the Orion A cloud (group $B_1$) is around 10 Myr old. The age is similar to the estimated age for the group related to the Orion A cloud (group C). However, the colour-magnitude diagram of group C (see Appendix B) shows that, not unexpectedly, many sources are brighter than the 10 Myr isochrone, and therefore likely younger.

A substantial luminosity spread has been observed in the colour-magnitude diagram of the stellar population towards the ONC \citep[see e.g.][]{ Jeffries2011, DaRio2010}. This spread represents the combined effect of a real age spread, possibly due to the presence of multiple populations \citep{Jerabkova2019,Beccari2017}, and of an apparent spread caused by other physical effects that scatter the measured luminosities, such as stellar variability and scattered light from circumstellar material. Age spreads are not included in our data modelling, therefore our age
estimate for group C should be considered as an upper estimate for the age of the stellar population towards the Orion A molecular cloud, which  also contains younger sources. The older population is more numerous than the younger ones, and therefore our age estimates are biased toward older ages.
The age estimate for group C and for all the other groups is very precise (see Table 2). This is partly an artefact of using a single isochrone set, and ignoring differential extinction as well as the effects mentioned above.
The presence of unresolved binaries in our data is also not taken into account, and could introduce biases towards younger age estimates, as unresolved binaries appear brighter than single stars. This could be the
case for example for groups B2 and B5 (see Fig. B). For the other groups the single star sequence is usually more numerous than the unresolved binary sequence, thus the fit results are weighted towards the single star
sequence.

In terms of age ranking, our age estimates agree with those found by \cite{Kounkel2018}: their Fig. 13 indicates indeed the presence of a diffuse older population, which however they find to be around 10 Myr old.
The difference in the maximum age they obtain is due to a number of differences in our fitting procedure: for example, they use $A_V = 0 \, \mathrm{mag}$ and a previous version of the \textit{Gaia} DR2 filters. 
Our results contradict instead what was found by \cite{Briceno2019}, who derive an age sequence that agrees with the long-standing picture of star formation starting in the 25 Ori region (also called Orion OB1a) and sequentially propagating towards the Belt region (Orion OB1b and 1c) and the Orion A molecular clouds (Orion OB1d).

\subsection{Sequential star formation and triggering in Orion}
The main result of this work is that the star formation history of the Orion region is complex and fragmentary. 
The Orion region is composed of many subgroups with different kinematic properties. Star formation started around 15 Myr ago \citep[or 20 according to ][]{Kos2018}, and still continues in the Orion A and B molecular clouds. The groups that we observe at the present time are sometimes spatially mixed (such as in the 25 Ori region) but their kinematics retain traces of their different origin.  
Figure \ref{fig:fig7} shows a schematic view of the Orion region, which summarises our results. The arrows represent the velocity vectors  \citep[in galactic Cartesian coordinates and corrected for the solar motion following][]{Schonrich2010} of the groups we identified, and are colour-coded by the group ages. The grey contours represent the stellar density integrated in the $Z$ (left), $Y$ (centre), and $X$ direction.  The Sun is at $(X, Y, Z) = (0, 0, 0) \, \mathrm{pc}$. 
\newline
\cite{Cantat-Gaudin2018} studied the Vela OB association, finding that a large fraction of the young stars in the region are not concentrated in clusters, but rather distributed in sparse structures, elongated along the Galactic plane. \cite{Krause2018} performed a multi-wavelength analysis of the Scorpius-Centaurus association, and suggested a refined scenario to explain the age sequence of the sub-groups that form the association. Similar to these studies, we find that the star formation history of Orion is not consistent with simple sequential star formation scenarios. Further, the traditional groups in which the Orion OB association is sub-divided are not monolithic episodes of star formation, but exhibit significant kinematic and physical sub-structure. 

We do not observe any clear age gradient nor any clear evidence of triggering in the kinematic properties of the groups \citep[such as those predicted for instance by][]{Hartmann2001}. 
As \cite{Cantat-Gaudin2018} suggest, the difference in velocity that are observed might be the result of galactic shear, or the consequence of a velocity pattern already imprinted in the filaments belonging to the parent molecular cloud these young populations formed from. The disposition in space of the clusters might reflect the structure of their parental molecular clouds: however this should be confirmed by specific simulations of the star formation process in the Orion region. 

\section{Conclusions}\label{sec:7}
In this work we study the 3D structure, the kinematics, and the age ordering of the young stellar groups of the Orion star forming region, making use of \textit{Gaia} DR2.
\begin{itemize}
\item We select young sources by applying simple cuts in the $M_G$ vs. $G_{BP} - G_{\mathrm{RP}}$ colour-magnitude diagram, and we study their density distribution in 3D galactic coordinates. 
\item We normalise our 3D density map between 0 and 1, and we select only the sources above a threshold of 0.5.
We then apply the DBSCAN clustering algorithm to identify groups in 3D space and we analyse their properties in terms of ages and kinematics.
\item We first inspect the proper motions of all the groups. In some cases we find that single groups in 3D space show sub-structures in their proper motion distribution. In this case we further sub-divide the groups, making simple cuts based on the proper motion distribution. We then apply a kinematic modelling code that we use to retrieve average motions, velocity dispersion, and isotropic expansion for all the groups identified. 
\item By comparing the 3D velocities of all the groups, we find evidence of kinematic sub-structures.
\item We compute ages and extinctions for all the groups by using a 2D maximum likelihood approach.We find that star formation in Orion started around 15 Myr ago in two groups, one towards the Belt region, and one towards the 25 Ori region. 
\item We do not find any clear age gradient, or any evidence of sequential star formation propagating from the 25 Ori region towards the Belt region and the Orion A and B molecular gas. 
\end{itemize}
\noindent
In conclusion, the picture of the Orion that we obtain from this study is that of a 
highly sub-structured ensemble of young stars with different ages, with several kinematic groups, mixed in 3D space and overlapping in the sky. These results do not agree well
with sequential star formation models, and would require designated specific simulations to be fully explained. 
\newline
The limited number of radial velocities available for most of the groups, as well as their large uncertainties, does not allow to characterise fully the internal kinematics of the clusters, or establish the presence of an-isotropic expansion. 
Future, ground based spectroscopic surveys could provide precise radial velocities for a large sample of sources, which, combined with the next \textit{Gaia} releases, will allow to better probe the internal kinematics of young clusters and OB associations. 

\begin{acknowledgements}
We thank the referee for their comments, which improved the manuscript.
This work has made use of data from the European Space Agency (ESA)
mission {\it Gaia} (\url{https://www.cosmos.esa.int/gaia}), processed by
the {\it Gaia} Data Processing and Analysis Consortium (DPAC;
\url{https://www.cosmos.esa.int/web/gaia/dpac/consortium}). Funding
for the DPAC has been provided by national institutions, in particular
the institutions participating in the {\it Gaia} Multilateral Agreement.
This project was developed in part at the 2018 NYC Gaia Sprint, hosted by the Center for Computational Astrophysics at the Simons Foundation in New York City. \\
This work has made extensive use of  Matplotlib \citep{matplotlib},  scikit-learn \citep{scikit-learn}, and TOPCAT \citep[\url{http://www.star.bris.ac.uk/~mbt/topcat/}]{topcat}. This work would have not been possible without the countless hours put in by members of the open-source community all around the world.   
\end{acknowledgements}

\begin{appendix}

\section{Testing the kinematic modelling code with simulated clusters}
We generate a sample of N = 200 stars which mimics the kinematics properties of young clusters and we test our code by changing a) the position of the sample (in particular its distance to the Sun), b) the velocity dispersion, and c) the expansion coefficient ($\kappa$) value. In particular we are interested in the ability of the code to retrieve the correct value for $\kappa$, especially when not all the radial velocities of the cluster members are provided.
\subsection{Simulation set up}
The simulated star positions are drawn from Gaussian distributions with $\sigma = 2 \, \mathrm{pc}$. 
The velocity of each simulated star is drawn following the same assumption as in L00, i.e. 
from a Gaussian distribution centred in $\bm{v_0}$ with a small velocity dispersion $\sigma$. We include expansion following Eq. 9, chosing $\kappa = 0.1 \, \mathrm{km \, s^{-1} \, pc^{-1}}$. 
\newline
We obtain the observed quantities (positions, parallax, proper motions, and radial velocities)\footnote{To do the transformation we make use of the \texttt{pygaia} routine \texttt{phaseSpaceToAstrometry}.} by adding typical \textit{Gaia} errors in the Orion region drawn from Gaussian distribution with widths $0.1 \, \mathrm{mas}$, $0.1 \, \mathrm{mas \, yr^{-1}}$,  and $3 \, \mathrm{km \, s^{-1}}$  respectively.
\subsection{Simple tests}
We simulate two clusters at different distances and with different velocities (see Tables 1 and 2, respectively): cluster A is similar in terms of kinematics $\bm{v_{0,I}} = (-5.0, 45.0, 6.0) \, \mathrm{km \, s^{-1}}$ and distance $(x_{0, I}, y_{0, I}, z_{0, I}) = (17.89,  42.14, 13.16) \, \mathrm{pc}$ to the Hyades cluster; cluster B is instead resembling the 25 Ori cluster: $(x_{0,I}, y_{0, I}, z_{0, I}) = (52.96, 343.97,  10.21) \, \mathrm{pc}$ and   $\bm{v_{0,I}} = (0.0, 20.0, 0.0) \, \mathrm{km \, s^{-1}}$ . We run the simulations in five different scenarios for both the simulated clusters:
\begin{enumerate}
    \item $\sigma_v = 0.3 \, \mathrm{km \, s^{-1}}$ and  $\kappa = 0.1 \, \mathrm{km \, s^{-1} \, pc^{-1}}$.
    \item $\sigma_v = 1.0 \, \mathrm{km \, s^{-1}}$ and  $\kappa = 0.1 \, \mathrm{km \, s^{-1} \, pc^{-1}}$;
    \item $\sigma_v = 0.3 \, \mathrm{km \, s^{-1}}$,  $\kappa = 0.1 \mathrm{km \, s^{-1} \, pc^{-1}}$, and a fraction $f = [10\%, 50\%, 95\%]$ of stars without measured radial velocities.
\end{enumerate}
The average velocities are always retrieved quite correctly in both cases; $\sigma$ and $\kappa$ are retrieved correctly for cluster A, however we notice that for cluster B the value of  $\kappa$ is usually underestimated, while  $\sigma$ is usually slightly over-estimated. 
When the number of observed radial velocities is too low, 
the expansion parameter can not be retrieved as it can not be separated from $\bm{v_0}$ from astrometric data only. In the cases when this happens, we do not give any estimate for the expansion term $\kappa$. When there are no radial velocities available the velocity is very poorly constrained,  especially for cluster B: in this case we do not give estimates for the velocities. 
When 10\% or 50\% of the measured radial velocities are missing,  the errors on the estimated parameters are of the same order of magnitude as in the other cases were all the kinematic data are available. However, not unexpectedly, when only 5\% of the radial velocities is available, the error on the $v_y$ parameter is roughly one order of magnitude larger than in the other cases. 
\subsection{Realistic tests}
In the real case it is likely that the clusters selected with the DBSCAN algorithm have both stars without measured radial velocities and kinematic outliers. We therefore further tested our code for cluster in two cases (see Table 3). In the first one we include 20 kinematic outliers in our simulated clusters: the kinematic outliers have a broader spatial distribution than the simulated cluster members ($\sigma = 5 \, \mathrm{pc}$), and their velocities are drawn from a Gaussian distribution with mean $20 \, \mathrm{km \, s^{-1}}$ in $x_I, y_I, z_I$, and dispersion $\sigma_v= 10 \, \mathrm{km \, s^{-1}}$. In the second one we include 20 kinematic outliers and we remove the 10\% of measured radial velocities.  In both cases, after the exclusion procedure the parameters are retrieved correctly. We notice that also in this case the expansion coefficient $\kappa$ is under-estimated (roughly by a factor of 2), while $\sigma_v$ is slightly
over-estimated.
\subsection{Initial conditions}
To test whether the initial conditions of the minimisation have an impact on the estimated parameters, we performed 100 runs with  initial guesses for the mean cluster velocity components, the velocity dispersion, and the expansion term $\kappa$ drawn randomly from a Gaussian distribution centred on the mean parameters, with dispersion equal to the 20\% of their real values.
\cite{Reino2018} performed similar tests on the Hyades cluster (which as said above is kinematically similar to our cluster A), finding essentially no dependence from the estimated parameters from the initial conditions. Thus, we repeat these tests only on our simulated cluster B.
\newline
\textit{B.1: $\sigma_v = 0.3 \, \mathrm{km \, s^{-1}}$ and  $\kappa = 0.1 \, \mathrm{km \, s^{-1} \, pc^{-1}}$}. We find that in general the minimisation results do not strongly depend on the initial parameters, however if the velocity dispersion $\sigma_v$ is over-estimated and/or the velocity in the $x_I$ component is under- or over-estimated then the velocity in the $y_I$ component is also under- or over-estimated.
\newline
\textit{B.2: $\sigma_v = 1. \, \mathrm{km \, s^{-1}}$ and  $\kappa = 0.1 \, \mathrm{km \, s^{-1} \, pc^{-1}}$}. We find that the minimisation results do not depend on the initial parameters in any case. This is reassuring, as the values for $\sigma_v$ in the clusters considered here are larger than $0.3 \, \mathrm{km \,  s^{-1}}$. In the cases with  $\sigma_v = 1. \, \mathrm{km \, s^{-1}}$ and missing radial velocities (for 20, 100, and 190 stars respectively), the estimated parameters are retrieved correctly for any choice of initial conditions, except for the expansion parameter $\kappa$, that is  underestimated. If outliers are present, the parameters are retrieved correctly after the exclusion procedure. 

\begin{table*}
  \begin{center}
    \caption{Results of the tests of the kinematic modelling for cluster A.}
    \label{tab:table1}
    \begin{tabular}{llllll}
    & $v_{x,I} \, [\mathrm{km \, s^{-1}}]$  & $v_{y,I} \, [\mathrm{km \, s^{-1}}]$ & $v_{z,I} \, [\mathrm{km \, s^{-1}}]$ &
       $\sigma_v  \, [\mathrm{km \, s^{-1}}]$ & $\kappa  \,  [\mathrm{km \, s^{-1} \, pc^{-1}}]$ \\
    
      \hline
      \hline
      Initial values & -5.0 &  45.0 & 6.0 & 0.3 & 0.1\\
      A.1     &-5.9 &  45.6 &  5.57 & 0.3 & 0.1 \\
	   	             & -5.88 $\pm$ 0.03 &  45.57 $\pm$ 0.05 & 5.56 $\pm$ 0.027 & 0.32 $\pm$ 0.01 & 0.1 $\pm$ 0.01\\
	 \hline
	 A.2       &-5.9 &  45.6 &  5.57 & 1.0 & 0.1 \\
	                  & -5.96 $\pm$ 0.08 &  45.5 $\pm$ 0.1 & 5.54 $\pm$ 0.08 & 1.01 $\pm$ 0.03 & 0.06 $\pm$ 0.02\\
	 \hline
	 
	 A.3   & &   &   &\\  
	 20/200 missing radial velocities & -5.873 $\pm$ 0.03 &  45.6 $\pm$ 0.06 &  5.586 $\pm$ 0.03 &  0.31 $\pm$ 0.01 & 0.1 $\pm$ 0.01  \\
	   	                    
	 100/200 missing radial velocities & -5.91 $\pm$ 0.035 &  45.55 $\pm$ 0.07 &  5.564 $\pm$ 0.03 &  0.3 $\pm$ 0.01 & 0.1 $\pm$ 0.1  \\
	   
	   190/200 missing radial velocities & -6.0 $\pm$ 0.4 &  45.0 $\pm$ 1.0 &  5.5 $\pm$ 0.3 &  0.3 $\pm$ 0.01 & 0.1 $\pm$ 0.03  \\
	  
    \end{tabular}
  \end{center}
\end{table*}

\begin{table*}
  \begin{center}
    \caption{Results of the tests of the kinematic modelling for cluster B.}
    \label{tab:table2}
    \begin{tabular}{llllll} 
       & $v_{x,I} \, [\mathrm{km \, s^{-1}}]$  & $v_{y,I} \, [\mathrm{km \, s^{-1}}]$ & $v_{z,I} \, [\mathrm{km \, s^{-1}}]$ &
       $\sigma_v  \, [\mathrm{km \, s^{-1}}]$ & $\kappa  \,  [\mathrm{km \, s^{-1} \, pc^{-1}}]$ \\
      \hline
      \hline
       Initial values & 0.0 & 20. & 0.0 & 0.3 & 0.1 \\
	   B.1    &  0.66 &  19.73 & 0.53 & 0.3 & 0.1 \\
	                      & 0.65 $\pm$ 0.03 &  19.83 $\pm$ 0.063 & 0.51  $\pm$ 0.026 & 0.36 $\pm$ 0.013 &  0.07 $\pm$ 0.01\\
	   \hline
	   B.2    &  0.66 &  19.73 & 0.53 & 1.0 & 0.1 \\                    
	                  &  0.65 $\pm$ 0.07 &  19.8 $\pm$ 0.1 & 0.56 $\pm$ 0.07 &  1.0 $\pm$ 0.03 & 0.05 $\pm$ 0.01 \\
	   \hline
	   B.3   & &   &   &\\ 
	   20/200 missing radial velocities  & 0.69 $\pm$ 0.03 &  19.80 $\pm$ 0.06 &  0.57 $\pm$ 0.02 &  0.33 $\pm$ 0.01 & 0.05 $\pm$ 0.006  \\
	   	                    
	   100/200 missing radial velocities  & 0.67 $\pm$ 0.03 &  19.87  $\pm$ 0.08 &  0.52 $\pm$ 0.03 &  0.38 $\pm$ 0.01 & 0.05 $\pm$ 0.007  \\
	   
	   190/200 missing radial velocities & 0.67 $\pm$ 0.043 &  19.89 $\pm$ 0.232 &  0.56 $\pm$ 0.025 &  0.32 $\pm$ 0.01 & 0.09 $\pm$ 0.008  \\
	  
    \end{tabular}
  \end{center}
\end{table*}

\begin{table*}
  \begin{center}
    \caption{Results of the tests of the kinematic modelling for cluster B, with missing radial velocities and outliers.}
    \label{tab:table3}
    \begin{tabular}{llllll} 
       & $v_{x,I} \, [\mathrm{km \, s^{-1}}]$  & $v_{y,I} \, [\mathrm{km \, s^{-1}}]$ & $v_{z,I} \, [\mathrm{km \, s^{-1}}]$ &
       $\sigma_v  \, [\mathrm{km \, s^{-1}}]$ & $\kappa  \, [\mathrm{km \, s^{-1} \, pc^{-1}}]$ \\
      \hline
      \hline
	   No missing radial velocities and 20 outliers & & & & & \\
	   \hline
	   Initial parameters & 0.0 & 20. & 0.0 & 0.3 & 0.1 \\
	   Real values &  0.66 &  19.73 & 0.53 & 0.3 & 0.1 \\
	   First iteration  &  0.62  $\pm$ 6. & 19.88 $\pm$ 6. & 0.71 $\pm$ 6 &  88.09 $\pm$ 2.4 &  -0.17 $\pm$ 0.4 \\
	   After exclusion procedure   &  0.63  $\pm$ 0.03 & 19.75 $\pm$ 0.06 & 0.52 $\pm$ 0.03 &  0.33 $\pm$ 0.01 &  0.05 $\pm$ 0.006 \\
	   
	   20 missing radial velocities and 20 outliers & & & & & \\
	   \hline
	   Initial parameters & 0.0 & 20. & 0.0 & 0.3 & 0.1 \\
	   Real values &  0.66 &  19.73 & 0.53 & 0.3 & 0.1 \\
	   First iteration  &  0.64  $\pm$ 4.85 & 19.88 $\pm$ 5.2 & 0.6 $\pm$ 4.85 &  72. $\pm$ 2. &  0.4 $\pm$ 0.3 \\
	   After exclusion procedure   &  0.68  $\pm$ 0.03 & 19.83 $\pm$ 0.06 & 0.55 $\pm$ 0.03 &  0.37 $\pm$ 0.01 &  0.07 $\pm$ 0.007 \\
    \end{tabular}
  \end{center}
\end{table*}

\section{Color magnitude diagrams}\label{appendixB}
\begin{figure*}\label{fig:appendixB}
\begin{center}
    \includegraphics[width = 0.8\hsize, height =0.95\vsize]{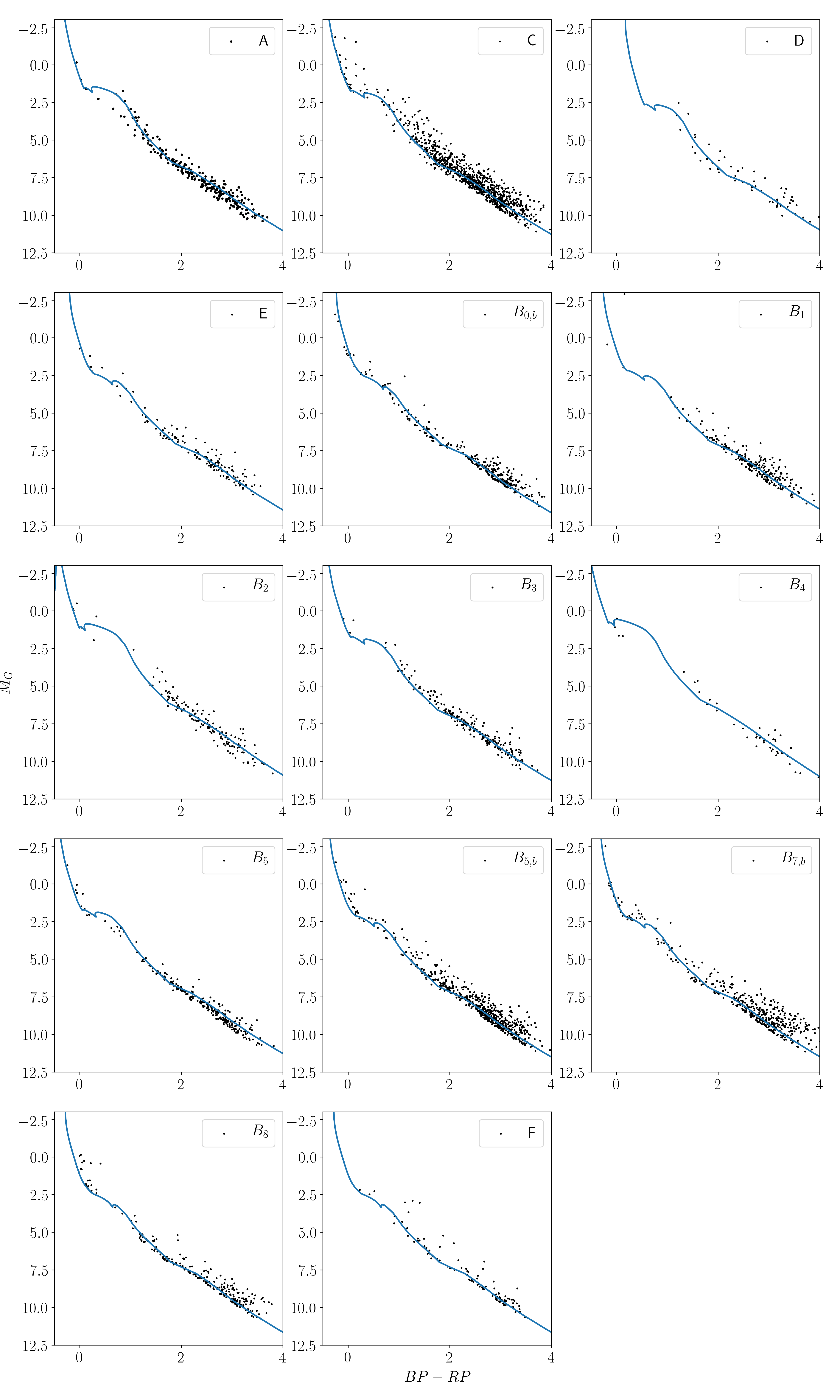}
    \caption{$M_G$ vs. $G_{BP}-G_{\mathrm{RP}}$ colour magnitude diagram for the groups selected in Section 4. The blue solid lines correspond to the best fitting isochrones, derived in Section 5.}
\end{center}
\end{figure*}

Fig. \ref{fig:appendixB} shows the colour magnitude diagram for the groups that we identified in Section 4.

\end{appendix}

\bibliographystyle{aa}
\bibliography{bibliografy}


\end{document}